
\catcode`\@=11 
\newbox\@tempboxa
\newdimen\@tempdima
\newdimen\fboxrule
\fboxrule=0.3pt
\newdimen\fboxsep
\fboxsep=5pt
\def\fbox#1{\leavevmode
        \setbox\@tempboxa\hbox{#1}\@tempdima
        \fboxrule \advance \@tempdima \fboxsep \advance \@tempdima \dp
        \@tempboxa \hbox {\lower \@tempdima \hbox {\vbox {\hrule height
        \fboxrule \hbox {\vrule width \fboxrule \hskip \fboxsep \vbox
        {\vskip \fboxsep \box \@tempboxa \vskip \fboxsep }\hskip \fboxsep
        \vrule width \fboxrule } \hrule height \fboxrule }}}}

\catcode`\@=12 

\input phyzzx
\hsize=40pc
%
%
%
%
%
%

\let\refmark=\NEWrefmark
\def\define#1#2\par{\def#1{\Ref#1{#2}\edef#1{\noexpand\refmark{#1}}}}

\define\hata H.~Hata, `Soft dilaton theorem in string field theory',
       {\it Prog.\ Theor.\ Phys.\ } {\bf 88}(1992):1197.

\define\hatanagoshi{H.~Hata and Y.~Nagoshi, `Dilaton and classical solutions
in pregeometrical string field theory', {\it Prog.\ Theor.\ Phys.\ }
{\bf 80}(1988):1088.}

\define\yoneya{T.~Yoneya, `String-coupling constant and dilaton vacuum
expectation value in string field theory',  Phys. Lett. B {\bf 197}(1987):76.}

\define\kugozwiebach{T.~Kugo and B.~Zwiebach, `Target space duality as a
symmetry of string field theory', {\it Prog.\ Theor.\ Phys.\ }
{\bf 87}(1992):801.}

\define\distlernelson{
J.~Distler and P.~Nelson, `Topological couplings and contact terms in
2-D field theory', Comm. Math. Phys. {\bf 138} (1991) 273;
`The Dilaton Equation in Semirigid String Theory',
{\it Nucl.\ Phys.\ } {\bf B366}(1991):255.}

\define\senzwiebachtwo{A.~Sen and B.~Zwiebach, `Quantum background
independence of closed string field theory', {\it Nucl.\ Phys.\ } {\bf
B423}(1994):580, hep-th/9311009.}

\define\campbell{M.~Campbell, P.~Nelson and E.~Wong, `Stress tensor
perturbations in conformal field theory', {\it Int.\ Jour.\ Mod.\
Phys.\ } {\bf A6}(1991):4909.}

\define\ranganathan{K.~Ranganathan, `Nearby CFT's in the operator
formalism: the role of a connection', {\it Nucl.\ Phys.\ } {\bf
B408}(1993):180. }

\define\bais{F.~A.~Bais, P.~Bouwknegt, M.~Surridge and K.~Schoutens,
`Coset construction for extended virasoro algebras', {\it Nucl.\
Phys.\ } {\bf B304}(1988):371.}

\define\leclair{A.~LeClair, M.~E.~Peskin and C.~R.~Preitschopf,
`String field theory on the conformal plane', {\it Nucl.\ Phys.\ }
{\bf B317}(1989):411.}

\define\senzwiebachtwo{A.~Sen and B.~Zwiebach, `Quantum background
independence of closed string field theory', {\it Nucl.\ Phys.\ } {\bf
B423}(1994):580, hep-th/9311009.}

\define\bergmanzwiebach{ O.Bergman and B.Zwiebach `The dilaton theorem
and closed string backgrounds' Nucl.~Phys. {\bf B441}(1995):76,
hep-th/9411047.}

\define\li{M.~Li, `Correlators of special states in $c=1$ Liouville
theory', {\it Nucl.\ Phys.\ } {\bf B382}(1992):242.}

\define\wittenzwiebach{E.~Witten and B.~Zwiebach, `Algebraic
structures and differential geometry in 2-D string theory'
{\it Nucl.\ Phys.\ } {\bf B377}(1992):55,  hep-th/9201056.}

\define\siegelzwiebach{W.~Siegel and B.~Zwiebach, `Gauge string
fields', {\it Nucl.\ Phys.\ } {\bf B263}(1986):105.}

\define\rahmanzwiebach{S.~Rahman and B.~Zwiebach, `Vacuum vertices and
the ghost-dilaton', submitted to {\it Nucl.\ Phys.\ }
B, hep-th/9507038.}

\define\polchinski{J. Polchinski, `Factorization of bosonic string amplitudes',
Nucl. Phys. {\bf B307} (1988) 61.}

\define\ademollo{J.~Shapiro, `On the renormalization of dual models',
{\it Phys.\ Rev.\ } {\bf D11}(1975):2937;\hfill\break M.~Ademollo, A.~D'Adda,
R.~D'Auria, F.~Gliozzi, E.~Napolitano, S.~Sciuto, and P.~Di~Vecchia,
`Soft dilatons and scale renormalization in dual theories', {\it
Nucl.\ Phys.\ } {\bf B94}(1975):221.}

\define\mende{P.~Mende, `Ghosts and the $c$-Theorem', {\it Phys.\
Rev.\ Lett.\ } {\bf 63}(1989):344.}

\define\distnel{ J.~Distler and P.~Nelson `New discrete states of
strings near a black hole', {\it Nucl.\ Phys.\ } {\bf B374}(1992):123.}

\define\astbel{A.~Astashkevish and A.~Belopolsky `String center of
mass operator and its effect on BRST cohomology' MIT-CTP 2471.}

\define\mahapatra{S.~Mahapatra, S.~Mukherji, and A.~M.~Sengupta, `Target
space interpretation of new moduli in 2D string theory',
{\it Mod.\ Phys.\ Lett.\ } {\bf A7}(1992):3119,  hep-th/9206111.}

\def\ve{\varepsilon}
\def\Lm{\Lambda}
\def\np#1{:\!#1\!:}           
\def\half{{1\over2}}

\def\a{\alpha}
\def\A{{\cal A}}
\def\B{{\cal B}}
\def\B{{\cal B}}

\def\D{{\cal D}}

\def\G{{\cal G}}

\def\H{\widehat{\cal H}}

\def\K{{\cal K}}
\def\L{{\cal L}}

\def\O{{\cal O}}

\def\Q{\hat{Q}}

\def\V{{\cal V}}
\def\V{{\cal V}}

\def\bra#1{\langle #1 |}
\def\braket#1#2{\langle#1|#2\rangle}

\def\ds{\displaystyle}

\def\d{{d}}
\def\e{{\epsilon}}

\def\frac#1#2{{#1\over#2}}

\def\inbar{\,\vrule height 1.5ex width .4pt depth0pt}\def\IC{\relax
\hbox{$\inbar\kern-.3em{\rm C}$}}
\def\ket#1{| #1 \rangle}
\def\ket#1{| #1 \rangle}
\def\k{{\kappa}}

\def\ov{\overline}

\def\p{\partial}

\overfullrule=0pt
\baselineskip 15pt plus 1pt minus 1pt
\nopubblock
\def\Q{Q^{\rm ext}}
\def\He{\H^{\rm ext}}         
\def\dl{\delta}
\def\Dg{{D_g}}                
\def\Dx{{\cal D}_X}           
\def\dx#1{{\Dx^{#1}}}           %
\def\dm#1{{D^{#1}}}           
\def\Xg{{\chi_g}}             
\def\xm#1{{\chi_m^{#1}}}        
\def\aux#1{\xi^{#1}}          
\def\inbar{\,\vrule height 1.5ex width .4pt depth0pt}
\def\QQ{{Q}}
\def\corr#1{\big\langle#1\big\rangle}

\def\met{\eta}                
{}~ \hfill \vbox{%
                 \hbox{MIT-CTP-2450}
\hbox{hep-th/9511077}\hbox{
} }\break
\title{WHO CHANGES THE STRING COUPLING ?}
\author{Alexander Belopolsky  \foot{E-mail address: belopols@marie.mit.edu }
and Barton Zwiebach \foot{E-mail address: zwiebach@irene.mit.edu.
\hfill\break Supported in part by D.O.E.
cooperative agreement DE-FC02-94ER40818.}}
\address{Center for Theoretical Physics,\break
Laboratory of Nuclear Science\break
and Department of Physics\break
Massachusetts Institute of Technology\break
Cambridge, Massachusetts 02139, U.S.A.}
\vfill
\centerline{\it Submitted to Nuclear Physics B}
\vfill
\abstract
{In general bosonic closed
string backgrounds the ghost-dilaton is not the only
state in the semi-relative BRST cohomology that can change the
dimensionless string coupling. This fact is used
to establish complete dilaton theorems in closed string field theory.
The ghost-dilaton, however, is the
crucial state: for backgrounds where
it becomes BRST trivial we prove that
the string coupling becomes an unobservable parameter of the string
action. For  backgrounds  where the matter CFT includes free uncompactified
bosons we introduce a refined BRST problem by including the zero-modes
$``x"$ of the bosons  as legal operators on the complex.
We argue that string field theory
can be defined on this enlarged complex and that
its BRST cohomology captures accurately the notion of a string background.
In this complex the ghost-dilaton appears to be the only BRST-physical state
changing the string coupling. }
\vfill
\endpage

\singlespace
\chapter{Introduction and Summary}

The soft-dilaton theorem is an old result in critical string
theory. It is stated as a property of string amplitudes for on-shell vertex
operators. A physical string amplitude involving a zero-momentum dilaton is
written in terms of derivatives, with respect to the
dimensionless coupling and
the slope parameter, of the string amplitude with the dilaton suppressed
[\ademollo]. It is  natural to ask what does this result tell us about
the role of the dilaton {\it field} in the string action. There has been
much work on the role of the dilaton in effective field theory limits
of strings. Our interest here is on the role of the dilaton in the complete
string action. This line of work  began with Yoneya [\yoneya]
who investigated the dilaton theorem using light cone string field theory.
Subsequent studies [\hatanagoshi,\kugozwiebach,\hata] considered the
dilaton in the context of covariantized light-cone string field theories.

A dilaton state has a component built by acting on the vacuum
by operators from the ghost sector. This component is called the ghost-dilaton
$\ket{D_g}$, and its relevance for the on-shell dilaton theorem was studied in
Ref.[\distlernelson]. This work was extended recently to prove the
off-shell ``ghost-dilaton theorem" in  covariant quantum
closed string field theory [\bergmanzwiebach,\rahmanzwiebach]. This result
states that an infinitesimal shift along the zero-momentum ghost-dilaton
changes the quantum string action, or more precisely, the
path integral string measure,
in a way equivalent to a shift in the dimensionless string coupling.
This work showed concretely that conformal field theories and string
backgrounds are not in  one to one correspondence: while the ghost-dilaton
deforms the string background, it does
not deform the conformal field theory underlying the string background. The
string background has a parameter which is absent in the conformal
field~theory.

A study of the ghost-dilaton alone is not enough
to understand how the value of the string coupling can be changed
in string theory. In critical string theory the zero-momentum
``matter-dilaton" also shifts the coupling constant.
Moreover, the properties of the ghost-dilaton depend on the matter sector of
the conformal theory.
In critical string theory it is a nontrivial BRST-physical state, but
in two-dimensional string theory, for example, it becomes  BRST-trivial.
If the ghost-dilaton is trivial one may think that the string coupling
cannot be changed. This is not correct, a shift of the ghost dilaton will
always shift the string coupling. What happens is that the string
coupling becomes unobservable. Thus the
ghost-dilaton plays a fundamental role: if it is trivial
the background has no string coupling constant parameter.
This is one of the main results of this paper.

Another point we develop in detail is the analysis of conformal
field theory deformations and string background deformations
induced by the dimension $(1,1)$ primary
field $\partial X\cdot\bar\partial X$. In critical string theory
this is the matter-dilaton.   We first consider the
analog of this state
in a conformal theory containing a single scalar field $X$
living on the real line.
We ask if $\partial X\bar\partial X$  deforms the conformal theory.
The deformation involves integrating the two-form $\partial X\bar\partial X
dz\wedge d\bar z$ over the surface, and this two form can be written
as the exterior derivative $[\, d\,( X\bar \partial X d\bar z )]$.
Stokes theorem
cannot be used directly because $( X\bar \partial X ) $ is not primary,
and therefore $( X\bar \partial X d\bar z )$ is not a true one-form.
Apart from a piece that
can be absorbed
by a redefinition of the basis of the conformal theory, we show that
the deformation amounts to a scaling  of  correlators with a
factor proportional to the Euler number of the underlying
surface. Since the conformal theory
has non-vanishing central charge,  correlators
depend on the scale factor of the metric and
the deformation simply corresponds to
 a variation of this scale factor.
Strictly speaking, the conformal field theory has
been deformed:  the correlators of the two theories on any {\it fixed}
surface cannot be made to agree with a redefinition of the basis of states.

If the matter conformal theory includes twenty six scalars and is
coupled to the ghost system, the operator $\partial X \cdot\bar\partial X$
does not deform the conformal theory. This result has been verified
earlier to various degrees of completeness in interesting works by
Mende [\mende] and Mahapatra {\it et.al.}[\mahapatra].
Indeed, following [\bergmanzwiebach]
the deformation can be absorbed by a change of basis
generated by the ghost-number operator.
For integrated correlators this cannot be done,
implying that  the matter dilaton
alters the string coupling.
We emphasize, however, that the matter dilaton does not change the
value of the dimensionful slope parameter $\alpha'$.

In addition to the dilaton we also consider the graviton trace ${\cal G}$.
This state, physical only at zero momentum, is
the linear combination of the ghost-dilaton
and the matter-dilaton that does not change the
string coupling.  The graviton trace can be written as $Q$ acting on
the state $\ket{\xi}$ created by
$(cX \cdot\partial X - \bar c X\cdot\bar\partial X)$. This state is usually
considered illegal since it
uses the operator $X$ which is not a scaling field of the conformal
theory. We argue in this paper that ${\cal G}$ is legally BRST trivial.
There is an immediate issue with this interpretation.
The failure of  ${\cal G}$ to decouple [\polchinski], easily verified in
$\langle {\cal G}, \hbox{phys},\hbox{phys} \rangle \not= 0$,
seems to be in contradiction with the
claim that ${\cal G}$ is BRST trivial: by contour deformation
the BRST operator that occurs in ${\cal G}$
can be made to act on the physical states giving zero.
We show that this is not a correct argument, the point being
that correlators involving $X$ are distributions. Careful use
of delta functions confirms  that
${\cal G}$ does not decouple.

Refining the discussion of Mahapatra {\it et.al.}\ [\mahapatra] and,
in agreement with them,
we claim that the  graviton trace ${\cal G}$ does not change any physical
property of the string background.
On the face of it this seems puzzling given the failure of decoupling for
${\cal G}$. There is no contradiction, however.
Strictly speaking,
a state leaves the physics of the background invariant
if it appears as the inhomogeneous term of a
nonlinear string field transformation that can be verified to
leave the string action invariant. A state
capable of having such behavior need not decouple. For ${\cal G}$
this is possible if correlators of operators including $X$'s
can be defined, and, once defined, obey the standard properties of
sewing and action of
the BRST operator. We discuss these matters and argue that
the requisite nonlinear field transformation is simply
a gauge transformation with gauge parameter $\ket{\xi}$.

Once we accept $X$
in the gauge parameters we must accept it in
the physical states as well.
Having a larger set of gauge parameters, we lose some physical states;
having a larger state space,  we may
also gain some new physical states.
We formalize this setup via
a new refined cohomology problem: BRST cohomology in the extended
(semirelative) complex where $X$ and powers of it are accepted as
legal operators.
The cohomology of this complex appears to capture accurately the
idea of a string background: we lose the states that do not change
the string background, as the
graviton trace, and we gain no states describing new physics. In this
cohomology the matter-dilaton is  the same as the ghost-dilaton.
Similarly, in two-dimensional string theory we show that
the states in the semirelative cohomology that are trivial in the
extended complex
do not appear to change the string background. A detailed computation
of BRST cohomology at various ghost numbers
in the extended complex of critical string theory
will be presented elsewhere [\astbel].

This paper is organized as follows. In section two we set up notation
and conventions. We discuss zero momentum physical states, and the
uses of the $X$ field operator. In section three we define the
extended BRST complex, explain the failure of ${\cal G}$ to decouple, and
argue that the usual BRST action on correlators holds in the extended complex.
We summarize
the properties of the new BRST cohomology following Ref.[\astbel].
In section four we  consider  CFT deformations induced by
the matter dilaton, and derive formulae for the integration of
insertions of matter dilatons over spaces of surfaces. We use these results
in section five to give a complete proof of the dilaton theorem in
closed string field theory. In section six we prove that whenever
the ghost-dilaton is BRST trivial the string coupling constant is
not observable. Finally, in section seven we illustrate some of our
work in the context of two-dimensional string theory.

\chapter{Zero momentum states and uses of
the ${\bf X^\mu (z, \bar z)}$ operator.}

In this section we begin by enumerating the zero-momentum
physical states in critical string theory. This enables us to
set our conventions and definitions for the dilaton, the ghost-dilaton,
the matter dilaton, and the graviton trace. We then elaborate on the
definition of the $X^\mu$ field operator and how it can be used
to write all zero momentum states, with the exception of the ghost-dilaton,
as BRST trivial states. The ghost-dilaton  can be also be
written in BRST trivial form, but, as is well known,
 in this case the gauge parameter
is not annihilated by $b_0^-$.  Finally, we discuss charges that can be
constructed using currents that involve the field operator $X^\mu$.

\section{Zero momentum physical states in critical string theory}

In this section we will list the ghost
number two physical states of critical bosonic closed strings
at zero momentum.  These states are defined as cohomology classes of
the  semirelative BRST complex.
We will find that only some of those states
can be obtained as a zero momentum limit
of  physical states that exist for non-zero momentum
(states corresponding to massless particles).
This is a familiar phenomenon  noted, for example, in Ref.[\distnel]
for the case of the fully relative closed string BRST cohomology.
The present section will also serve the purpose of setting up
definitions and conventions.

Let $\ket{\Psi}$ be the string field state and $Q$
the BRST operator. Physical states are defined by
$$Q\ket{\Psi}=0 \,, \eqn\brstt$$
up to gauge transformations
$$
\dl\,\ket{\Psi}=Q\ket{\Lm}.
\eqn\brst
$$
Here both $\ket{\Psi}$ and $\ket{\Lambda}$
must be annihilated by $b_0^-= b_0 -\bar b_0$.
To look for states that can be physical at zero momentum the relevant part
of the string field is
$$\eqalign{
\ket{\Psi} &=\,E_{\mu\nu}\, c_1\a_{-1}^\mu\ov c_1\ov\a_{-1}^\nu\ket{p} \cr
&\quad -  \ov A_\mu c_0^+c_1\a_{-1}^\mu\ket{p}+
  A_\mu c_0^+\ov c_1\ov\a_{-1}^\mu\ket{p}
\cr & \quad +
  F\, c_1c_{-1}\ket{p}-\ov F\, \ov c_1\ov c_{-1}\ket{p}+\cdots \cr }\eqn\fld$$
where $c_0^+=(1/2)(c_0+\ov c_0)$, and that of the gauge parameter
$$
\ket{\Lm}=\ve_\mu c_1\a_{-1}^\mu\ket{p}-
\ov\ve_\mu \ov c_1\ov\a_{-1}^\mu\ket{p}+
\ve c_0^+\ket{p}+\cdots\;.
\eqn\gge
$$
 Let us work at zero momentum $p_\mu = 0$.
Equation \brstt\ gives
 $A_\mu = \ov A_\mu =0$ and Eq.\brst\ gives us the gauge transformations
$\delta F = - \delta \ov F = -\phi$. It follows that at zero momentum
the $d^2$ degrees of freedom of $E_{\mu\nu}$ are unconstrained,
the combination $F+\ov F$ is gauge invariant and unconstrained,
and $F-\ov F$ can be gauged away. This gives a total of $d^2+1$
nontrivial BRST physical states in the semirelative complex.
For $p_\mu \not= 0$, it is well known that
there are $(d-2)^2$ nontrivial BRST states for each value of
momentum satisfying $p^2=0$ . These considerations indicate
that we have $(4d-3)$ states that are only physical at zero momentum.
These states are called discrete states.

The ($d^2+1$) zero-momentum physical states correspond to the
following CFT fields
$$\eqalign{
   \Dg&\equiv\half\,( c\p^2c-\ov c{\ov\p}^2\ov c)\,,\cr
\dm{\mu\nu}&\equiv c\ov c \p X^{\mu}\ov\p X^{\nu}.\cr}\eqn\Dmn$$
The  state associated to $\Dg$ is called the
ghost-dilaton and we will refer to the state associated to the
trace $\eta_{\mu\nu}
\dm{\mu\nu}$ as
the matter dilaton. In addition we identify
two relevant linear combinations of the ghost and the matter dilaton.
The first combination is the zero-momentum dilaton
$$D\equiv\met_{\mu\nu}\dm{\mu\nu}-\Dg\, ,\eqn\dilaton$$
which is the zero-momentum limit of the scalar massless state called the
dilaton. It is  recognized as such because the corresponding
spacetime field transforms as a scalar
under gauge transformations representing diffeomorphisms.
The second state is the ``graviton
trace"
$${\cal G}\equiv\met_{\mu\nu}\dm{\mu\nu}-{d\over2}\Dg\, .\eqn\gravtr$$
This state corresponds to the trace of the graviton field
in the convention where the
gravity action is of the form
$\int \sqrt g R dx$ without a factor involving the dilaton
(see, for example, Ref.[\siegelzwiebach]).

It is of interest to consider the gauge transformation generated by
$\ket{\Lambda}$ for the
case when we set $\ve =0$. We find
$$
\eqalign{\dl E_{\mu\nu}&=p_\mu\ov\ve_\nu+p_\mu\ve_\mu, \cr
         \dl F&=-p^\mu\e_\mu,\cr
         \dl\ov F&=-p^\mu\ov\ve_\mu.}
\eqn\gtrr
$$
Transforming  to coordinate space we obtain the linearized
gauge transformations
$$
\eqalign{
\dl E_{\mu\nu}(x) &= \p_\mu \ve_\nu(x) + \p_\nu \ov \ve_\mu(x)\,,\cr
\dl F(x) &=-\p^\mu\ve_\mu\,,\cr
\dl \ov F(x) & =-\p^\mu\ov\ve_\mu\,.\cr}\eqn\ngauge
$$
Now consider the  gauge parameters
$$\ve_\nu(x) = C_{\mu\nu}x^\mu,\quad
\ov\ve_\mu(x) = C_{\mu\nu}x^\nu,
\eqn\gauge$$
where $C_{\mu\nu}$ is a matrix of constants. Equation \ngauge\
implies that the following
constant field configurations are pure gauge:
$$ E_{\mu\nu}(x) = 2 C_{\mu\nu}\,, \quad
 F(x)=\ov F(x)=C_\mu^\mu\,\,.\eqn\fields$$
In
string field theory the coefficient $E_{\mu\nu}(p)$ in Eq.\fld\
should be interpreted as a Fourier component of the space-time field
$E_{\mu\nu}(x)$. The above spacetime constant field configurations
must correspond to zero momentum states.
The corresponding string field in \fld\ should be expected to be BRST
trivial. This requires that
$$\dm{\mu\nu}-{1\over2}\,\met^{\mu\nu}\Dg
\,\equiv\,\G^{\mu\nu}\,,\eqn\gravdef$$
is BRST trivial
$$ \G^{\mu\nu}=-\{\, \QQ\, ,\,\aux{\mu\nu}\, \}. \eqn\newgauge$$
In the ordinary closed string BRST complex there is no state
$\aux{\mu\nu}$ satisfying \newgauge. It is necessary to extend the
BRST complex to include  states  corresponding to
field configurations growing linearly in space-time. An example
is the state $\lim_{z\to 0}X^\mu(z,\ov z)\ket{0}$, which requires
the consideration of  $X^\mu(z,\ov z)$
as a field operator. We analyze this next.

\section{Definition of the $X^\mu$ operator}
In the ordinary CFT of $26$ free bosons only derivatives of
$X^\mu(z,\ov z)$ appear as conformal fields. These derivatives have
the  mode expansions
$$ i\,\p X^\mu(z, \ov z) = \sum_{n=-\infty}^\infty{\a^\mu_n\over z^{n+1}}\,,
\qquad
 i\,\ov\p X^\mu(z, \ov z) =
    \sum_{n=-\infty}^\infty{{\ov\a}^\mu_n\over{\ov z}^{n+1}}\,,\eqn\parX$$
where the $\a^\mu_n$'s are  operators with  commutation relations
$$[\a^{\mu}_m\, ,\a^{\nu}_n]=m\,\met^{\mu\nu}\delta_{n,m}\,\,,\qquad
[\ov\a^{\mu}_m\,
,\ov\a^{\nu}_n]=m\,\met^{\mu\nu}\delta_{n,m}\,\,.\eqn\commute$$
Formally integrating Eq.~\parX\ we  find
$$X^{\mu}(z,\ov{z}) = X^{\mu}_0-2\,i\,\a^{\mu}_0\log\,{|z|}+
\sum_{n\neq0}{i\,\a^{\mu}_n\over n z^n}
+\sum_{n\neq0}{i\,{\ov\a^{\mu}}_n\over {n \ov{z}}^n}\,\,.
\eqn\X$$
The zero mode operator $X^\mu_0$, which appears in \X\ as a constant of
integration is not specified by \parX\ and has to be defined
independently. It is standard to
interpret it as the position operator for the center of mass, and
taken to commute with all $\a$'s except the momentum operator $\a^{\mu}_0$
$$[X^{\mu}_0,\a^{\nu}_0] = i\,\met^{\mu\nu}. \eqn\pq$$
This interpretation is usually justified by canonical analysis of
the two-dimensional quantum field theory.
Eqs.~\X\ and~\pq\ provide
an abstract definition of $X^\mu(z,\ov{z})$ as an element of a Lie
algebra.

\section{A gauge parameter involving $X^\mu(z,\bar z)$}

Now we will try to use the $X^\mu$ field to find a gauge parameter which
will generate $\G^{\mu\nu}$. Equations \gge\ and \gauge\ suggest that
the proper candidate is
$$\aux{\mu\nu} =\half \,(c \np{ X^{\nu}   \p X^{\mu}} -
                         \ov c \np {X^{\mu}\ov\p X^{\nu}})\, .\eqn\xichi$$
We have
to explain what the products like  $\np{X^{\mu}\p X^{\nu}}$ or
$\np {X^{\nu}\ov\p X^{\mu}}$  mean. Normal ordering
amounts to placing annihilation operators to the right
of creation operators. In our
case it is not clear how to order the product of $X^{\mu}_0$ and
$\a_0$.  We adopt the following definition
$$\np{X^\nu\p X^\mu}(z,\ov{z}) \equiv \oint_z \, {\,R\,(\,\p
X^\mu(w)X^\nu(z,\ov{z})
\, )\over
w-z}{\d w\over 2\pi i}\,\,, \eqn\npdef$$
where $R$ denotes the necessary radial ordering.
Note that the integral is  contour independent because $\p X^\mu$
is a holomorphic field. As usual, to
evaluate the integral we replace the contour around $z$ by two
constant radius contours,
one with $|w| > |z|$, and the other with $|w| < |z|$.
Because of
radial ordering one must use different expansions for
$1/(w-z)$ in the two contours. A small calculation gives
$$\np{X^\mu\p X^\nu}(z, \ov{z}) = X^\mu(z,\ov{z})
                      \sum_{n\geq0}{-i\,\a_n^{\nu}\over z^{n+1}}
                    + \sum_{n<0} {-i\,\a_n^{\nu}\over z^{n+1}}
                       X^\mu(z,\ov{z})\,,\eqn\xx$$
which shows that the momentum zero mode $\a^\mu_0$ appears to the
right of the coordinate zero mode $X^\mu_0$.

We now calculate the action of the BRST operator on $\aux{\mu\nu}$:
$$
\eqalign{
\{\QQ,\aux{\mu\nu}(z,\ov{z})\}=&\oint \Bigl( T_m(z)c(z)+
   \np{c(z)\p c(z)b(z)}\Bigr)\,\aux{\mu\nu}(z,\ov{z})\,{\d z\over 2\pi i}
\cr
+&\oint \Bigl(\ov T_m(\ov z)\ov c(\ov z)+
   \np{ \ov c(\ov z)\ov\p\ov c(\ov z) b(\ov z)}\Bigr)
\,\aux{\mu\nu}(z,\ov{z})\,{\d \ov z\over 2\pi i}.}
$$
Using  Wick's theorem\foot{As explained in Ref.[\bais] Wick's theorem
for composite operators is valid when normal ordering is defined
as in \npdef.} to expand the operator product under the
integral, we obtain
$$\{\QQ\,,\,\aux{\mu\nu}\} =
     {1\over4}\,\met^{\mu\nu}(c\p^2c-\ov c{\ov\p}^2\ov c)
                - c\ov c \p X^{\mu}\ov\p X^{\nu}\,,\eqn\qaux$$
and, as expected,
$$\G^{\mu\nu}=-\{\QQ,\aux{\mu\nu}\}\, . \eqn\newgauge$$

The ghost dilaton, the first state in \Dmn,
is a nontrivial state in the
semirelative cohomology, but in the  absolute complex
it can be represented as [\distlernelson]
$$\Dg = -\,\{\QQ\,,  \Xg\}\,, \eqn\Dghost$$
where $\Xg$ is given by
$$ \Xg = -\half(\p c-\ov\p\ov c)\, .\eqn\Xgdef$$
The state $\Xg$ does not belong to $\H$ since it is not annihilated by
$b_0^-$. Using $\Xg$ we can represent
the matter states of \Dmn\ as $\{\QQ,\,\cdot\,\}$. Indeed, let
$$\chi^{\mu\nu}\equiv \aux{\mu\nu}+{1\over 2}\met^{\mu\nu}\,\Xg\,.\eqn\ximn$$
Unlike $\Xg$ and $\aux{\mu\nu}$, the state $\xm{\mu\nu}$ can be verified to
be a $(1,1)$ primary.  Now combining
Eq.~\Dghost\ with Eq.~\qaux\ we obtain
$$\dm{\mu\nu} =-\{\QQ, \chi^{\mu\nu}\},\eqn\Qximn$$
and conclude that $\chi^{\mu\nu}$ is the gauge parameter generating the
matter states of \Dmn.
We therefore conclude that all semirelative cohomology states at
zero momentum can be represented as $\{\QQ,\,\cdot\,\}$ when we allow
gauge parameters using the $X$ operator and/or violating the
$b_0^-=0$ condition. This information is summarized in the following table.

$$\hbox{\vbox{\offinterlineskip
\def\strut{\hbox{\vrule height 15pt depth 10pt width 0pt}}
\hrule
\halign{
\strut\vrule#\tabskip 0.1in&
\hfil$#$\hfil &
\vrule#&
\hfil$#$\hfil &
\vrule#&
\hfil$#$\hfil &
\vrule#\tabskip 0.0in\cr
&D_g={1\over2}(c\p^2c-\ov c\ov\p^2 c) &&D_g=-\{Q,\chi_g\} &&
\chi_g=-{1\over2}(\p c -\ov\p\ov c) & \cr
\noalign{\hrule}
&D^{\mu\nu}=c\ov c\p X^\mu\ov\p X^\nu &&D^{\mu\nu}=-\{Q,\chi^{\mu\nu}\} &&
\chi^{\mu\nu}=\xi^{\mu\nu}+{1\over2}\eta^{\mu\nu}\chi_g & \cr
\noalign{\hrule}
&{\cal G}^{\mu\nu}=D^{\mu\nu}-{1\over2}\eta^{\mu\nu}D_g &
&{\cal G}^{\mu\nu}=-\{Q,\xi^{\mu\nu}\} &&
\xi^{\mu\nu}={1\over2}(cX^\nu\p X^\mu-\ov cX^\mu\ov\p X^\nu)& \cr
\noalign{\hrule}
&\eqalign{D=&D^\mu_\mu-D_g\cr
           =&\textstyle{\cal G}^\mu_\mu+\left({d-2\over2}\right)D_g}&
&D_g=-\{Q,\chi_{{}_D}\} &
&\eqalign{\chi_{{}_D}=&\,\chi^\mu_\mu-\chi_g\cr
           =&\,\,\textstyle{\xi}^\mu_\mu+\left({d-2\over2}\right)\chi_g} & \cr
\noalign{\hrule}}}}$$

\section{Properties of
         $\np{X^{\nu}\p{}X^{\mu}}\d z$}

It is of interest to consider some additional properties
of  $\np{X^{\nu}\p{}X^{\mu}}\, \d z$ and
$\np{X^{\mu}\ov\p{}X^{\nu}}\,\d\ov{z}$. Both of
them have an exterior derivative proportional to the two-form
$\p X^\mu\ov\p X^\nu\, dz\wedge d\bar z$
$$- \d \,\Big(\!\np{X^{\nu}\p{}X^{\mu}}\d z\Big) =
\d\, \Big(\!\np{X^{\mu}\ov\p{}X^{\nu}}\d\ov{z}\Big)=
\p X^\mu\ov\p X^\nu\d z\wedge\d\ov z.
\eqn\dfr
$$
Nevertheless
$\np{X^{\nu}\p{}X^{\mu}}\,\d z$ is not a one
form because the field $\np{X^{\nu}\p{}X^{\mu}}$ is not
primary
$$
T(z)\np{ X^{\nu}\p X^{\mu}} (w,\ov{w}) = {-\met^{\mu\nu}\over (z-w)^3} +
         {\np{X^{\nu}\p X^{\mu}}(w,\ov{w})\over(z-w)^2} +
         {\p(\np{ X^{\nu}\p X^{\mu}})(w,\ov{w})\over z-w} +\cdots\;\,.
\eqn\txx$$
As a consequence we have an anomalous
transformation law under analytic maps
$$
\np{X^{\nu}\p X^{\mu}}(z',\ov z')\d z' =\left( \np{X^{\nu}\p X^{\mu}}
(z,\bar z) -\half\met^{\mu\nu} {\d^2 z/\d z'^2\over (\d z/\d z')^2}\right)\d z.
\eqn\trans
$$
See~[\leclair] for details. Similar results hold for
$\np{X^{\mu}\ov\p{}X^{\nu}}\d\ov{z}$.

This implies that
$\np{X^{\nu}\p{}X^{\mu}}\d z$ and/or
$\np{X^{\mu}\ov\p{}X^{\nu}}\d\ov{z}~$
cannot be integrated unambiguously over a contour unless
we fix a coordinate in the vicinity of it. Nevertheless, we can show
that an integral over a contractible path does not change if we make a
coordinate transformation which is holomorphic inside the
path. Indeed, according to Eq.~\trans, when we make a holomorphic
change  of coordinate the anomalous piece which appears in the
transformation law is a holomorphic (or antiholomorphic) one-form
whose integral over a contractible path is zero. Thus the
integrals
$\oint_{\gamma} \np{X^{\nu}\p X^{\mu}}\d z$
and $\oint_{\gamma} \np{X^{\mu}\ov\p X^{\nu}}\d\ov z\,$
are well defined if $\gamma$ is contractible  but they still depend on
the choice of the contour $\gamma$ itself. For later use
we define
$$\dx{\mu\nu}=-\oint_{|z| =1}{\np{X^{\nu}\p X^{\mu}}}(z,\ov z){\d z\over2\pi i}
=\oint_{|z| =1} \np{X^{\mu}\ov\p X^{\nu}}(z, \ov z){\d\ov z\over2\pi i}\,\,,
\eqn\dxints$$
where the equality follows by use of \dfr.  Using Eq.~\xx\
 we rewrite the above  as
$$\dx{\mu\nu}= -\oint_{|z| =1}{\d z\over2\pi i}\, X^\nu(z,\ov{z}) \sum_{n\geq0}
{-i\,\a_n^{\mu}\over z^{n+1}} -\oint_{|z| =1}{\d z\over2\pi i}\,
\sum_{n<0} {-i\,\a_n^{\mu}\over z^{n+1}}\,
 X^\nu(z,\ov{z})\,\,\, .\eqn\dxc$$
The integration is done using Eq.~\X.
Since we  integrate over the unit circle the logarithm
in \X\ vanishes, and  we obtain
$$\dx{\mu\nu} =  i X^{\nu}_0\a^{\mu}_0 - \sum_{n\neq0} \,{1\over n} \,
\a^{\mu}_n\,\ov\a^{\nu}_n\,.\eqn\dxres$$
This operator appeared earlier in the dilaton theorem analysis of
refs.[\yoneya,\hatanagoshi].

\chapter{Extended BRST complex}

In this section we  define an extended BRST complex where the
coordinate zero mode $X_0$
acts as a linear operator. We will show how the
BRST operator $\Q$ acts on this complex and
explain why BRST exact
states in this complex may not  decouple from
physical correlations. We will explain, through an example
why the usual BRST action on correlators holds in the extended
complex. This, together with the fact that sewing also holds
in the extended complex, implies that string field theory is
well defined in the extended complex. The explicit computation of the BRST
cohomology in this complex will be given in Ref.[\astbel]. The present
section  concludes with a review of the results of this computation.

\section{Definition of the Extended Complex}

We define the extended space of states at any
given momentum $p$ as a tensor product of the original state space
$\H_p$ with the space of polynomials of $D$ variables $x^\mu$:
$$\He_p = \IC\,[x^{\mu}]\otimes\H_p\, . \eqn\Hedef$$
A state in this complex is written in the form $P\otimes v$
where $P\in \IC\,[x^{\mu}]$ is
a polynomial in $x^\mu$, and $v\in\H_p$ is a vector from the
original state space. The operator $X^{\mu}_0$ acts by multiplying the
polynomial by $x^\mu$.
All the mode operators,  except for $\a_0^\mu$, act as they
acted on $\H_p$.  For $\a_0^\mu$ it is natural to define
$$\a_0^\mu\; P\otimes v =
   p^\mu P\otimes v -
   i\met^{\mu\nu}{\p P\over\p x^\nu}\otimes v\, ,\eqn\ao$$
preserving in this way the commutation relation
$[X_0^\mu , \alpha_0^\nu\,] = i\eta^{\mu\nu}$.
With the matter Virasoro generators written as
$$
  L_n={1\over2}\sum_m\met_{\mu\nu}\,\np{\a^\mu_m\,\a^\nu_{n-m}}\, ,
\qquad
\ov L_n={1\over2}\sum_m\met_{\mu\nu}\,\np{\ov\a^\mu_m\,\ov\a^\nu_{n-m}}\,,
\eqn\vir
$$
the BRST charge reads
$$
  \Q=\sum_n c_n L_{-n} -
  {1\over2}\sum_{m,n}(m-n)\np{c_{-m}c_{-n}b_{m+n}}+\,\hbox{a.h.}
\eqn\brst
$$
Although equations \vir\ and \brst\ have the standard form, when acting
on the extended complex we must use Eq.~\ao.
For the BRST operator this implies that
$$
 \Q\;  P\otimes v =
  P\otimes Q v-i{\p P\over\p x^\mu}\otimes \sum_{n=-\infty}^{\infty}
  (c_n\a^\mu_{-n}+\ov c_n\ov \a^\mu_{-n})v -
  \met^{\mu\nu}{\p^2 P\over\p x^\mu\p x^\nu}\otimes c_0^+v.
\eqn\qacts.
$$

\section{The failure of BRST decoupling}

The analysis of section two shows that the  states
$\G^{\mu\nu}$
are BRST trivial in the extended complex. Indeed, as written in the table
at the end of section 2.3,
$\G^{\mu\nu}=\{\QQ\, ,\, \xi^{\mu\nu}\, \}$, where $\xi^{\mu\nu}$
contains an explicit $X$ operator. Correlators involving an explicit
$X$ operator should be evaluated using
$X^\mu(z)=-i\left.{\p \over\p p_\mu}\exp(ip X(z))\right|_{p=0}$, which
operationally  means evaluating the correlator with $X^\mu(z)$ replaced
by $\exp(ip X(z))$ and evaluating the derivative $-i{\p \over\p p_\mu}$
of the resulting correlator at  $p=0$.

In general, the correlation functions of a BRST trivial state
with physical states are known to vanish. Indeed, let $\ket\phi=-Q\ket\chi$ be
BRST trivial, and  $\ket{\psi_k}$ for $k=1,\dots,n,$ be BRST physical
($\QQ\ket{\psi_k}=0$). It follows by contour deformation
that the BRST operator can be taken to act on the physical states giving
$$\corr{\phi\,\psi_1\cdots\psi_n}=\sum_{k=1}^n
         \corr{\chi\,\psi_1\cdots Q\psi_k\cdots\psi_n}=0.\eqn\naive$$
On the other hand, consider the three point function of
the BRST trivial zero-momentum state $\G^{\mu\nu}$
with two tachyons $\tau_p(z,\ov z)=c\ov c\exp(ipX)(z,\ov z)$. One readily
verifies that for $\mu\not= \nu$
$$\corr{\G^{\mu\nu}(z_1,\ov z_1)\,\tau_p(z_2,\ov z_2)\, \tau_q(z_3,\ov z_3)}=
  2\, p^\mu p^\nu\, |z_2-z_3|^{4-2p^2}(2\pi)^d\dl^d(p+q)\, ,\eqn\exmpl$$
and observes that this  is not zero for on-shell tachyons ($p^2=q^2=2$),
 in apparent contradiction \naive.
Since the  computation leading to \exmpl\ is beyond doubt we
must find why the usual argument for decoupling fails.

The  problem with \naive\ is
that when $\phi =\G^{\mu\nu}$, the field $X^\mu$ is present in
the  correlators under the sum, and we cannot make
sense of their on-shell values.
This is because  correlators are not functions but
rather distributions. In the
absence of $X$ fields, however, we associate an ordinary function to a
correlator because every correlator can be represented as an ordinary
function of momenta times the standard momentum conserving $\dl$-distribution
$$\corr{\psi_1\psi_2\cdots\psi_n}=
F\,(p_1,p_2,\dots,p_n)\,(2\pi)^d
\delta^d\,(p_1+p_2+\cdots+p_n)\, .\eqn\assf$$
When we say that the correlator $\corr{\psi_1\psi_1\cdots\psi_n}$
vanishes for particular values of momenta what we
really mean is that the function $F$ vanishes. On the
other hand, if the fields $\psi_i$  contain $X$ without derivatives,
correlators have a more general structure
$$\eqalign{
\corr{\psi_1\psi_2\cdots\psi_n}=&\,
\,F\,(p_1,\dots,p_n)\,\,\,(2\pi)^d\,\dl^d\,(p_1+\cdots+p_n)\cr
+&F^\mu(p_1,\dots,p_n)\,
\,\,(2\pi)^d\, \dl^d_{\;,\mu}(p_1+\cdots+p_n)+\cdots\,.\cr}\eqn\assfx$$
where  $\delta^d_{,\mu}(p) \equiv {d\over dp^\mu} \delta^d (p)$,
and the dots indicate possible terms with higher derivatives
of delta functions.
A function $F^\mu$ that vanishes on-shell can contribute to the
correlator if its derivative does not vanish on shell.
This implies that the right hand
side of Eq.~\naive\ need not vanish when the correlator contains an $X$.
We will illustrate this with an example. Rather than using the field
$\G^{\mu\nu}$, the point can be made by considering  open string theory
where the analogous field is the zero-momentum photon
$c\p X^\mu =\{Q,X^\mu\}$.  We therefore
examine the correlator of this state with two tachyons
$$
\corr{c\p X^\mu(x)\,ce^{ip_1X(x_1)}\,ce^{ip_2X(x_2)}}=
-ip^\mu_1\, |x_2-x_1|^{2-p_1^2}\, (2\pi)^d \, \dl^d(p_1+p_2)\,,\eqn\firstcm$$
where  $x, x_1, x_2$ denote the insertion points on the real axis.
On the other hand, using the BRST property we are led to write
$$\eqalign{
\corr{c\p X^\mu(x)\,ce^{ip_1X(x_1)}\,ce^{ip_2X(x_2)}}=&
-\corr{X^\mu(x)\,\{Q,ce^{ip_1X(x_1)}\}\,ce^{ip_2X(x_2)}}\cr
&+\corr{X^\mu(x)\,ce^{ip_1X(x_1)}\,\{Q,ce^{ip_2X(x_2)}\}} \,.\cr  }
\eqn\exmpl$$
Using $
\{Q,ce^{ipX(z)}\}=({p^2\over2}-1) c\p ce^{ipX(z)}$, we obtain
$$(I) = \corr{X^\mu(x)\,\{Q,ce^{ip_1X(x_1)}\}\,ce^{ip_2X(x_2)}}=
\Bigl({p_1^2\over2}-1\Bigr)
|x_1-x_2|^2\corr{X^\mu(x)\,e^{ip_1X(x_1)}\,e^{ip_2X(x_2)}}\,, \eqn\midw$$
and the remaining matter correlator is evaluated using the prescription
given at the beginning of the section
$$(I) =-i\,
\Bigl({p_1^2\over2}-1\Bigr)|x_1-x_2|^{2+p_1p_2}
\bigg[p_1^\mu\log {|x-x_1|\over |x-x_2|}\,(2\pi)^d\,\dl^d(p_1+p_2)
+(2\pi)^d\,\delta^d_{,\mu}(p_1+p_2)\bigg]\,.\eqn\jis$$
In this expression, the term including the ordinary delta function
is unchanged under
the simultaneous exchanges  $x_1\leftrightarrow x_2$ and
$p_1\leftrightarrow p_2$, and thus back in \exmpl\ we obtain
$$\eqalign{
\corr{c\p X^\mu(x)\,ce^{ip_1X(x_1)}\,ce^{ip_2X(x_2)}}&=
{i\over 2}\, (p_1^2-p_2^2)|x_1-x_2|^{2+p_1p_2}\,(2\pi)^d\,
\dl^d_{,\mu}(p_1+p_2)\,,\cr
&= {i\over 2}\, p\cdot q\,|x_1-x_2|^{2+(p^2-q^2)/4}\,(2\pi)^d\,
\dl^d_{,\mu}(p)\,,
\cr}\eqn\hskdi$$
where $p^\mu=p_1^\mu+p_2^\mu$ and $q^\mu=p_1^\mu-p_2^\mu$. Since
$xf(x)\dl'(x)= -f(0)\dl(x)$ we find
$$
\corr{c\p X^\mu(x)\,ce^{ip_1X(x_1)}\,ce^{ip_2X(x_2)}}=
-i \,p_1^\mu |x_1-x_2|^{2-p_1^2} (2\pi)^d \, \dl^d(p_1+p_2)\,, \eqn\uhgg$$
in agreement with \firstcm, and confirms the failure
of decoupling. This example also illustrates that
with proper treatment of
distributions the BRST property of
correlators holds in the extended complex. There is no problem with
the contour deformation arguments that are used to prove the BRST property.
This will also be the case when we deal with integrated correlators.
The sewing property of correlators is also preserved in the extended complex.
This follows from the definition of correlators when a field $X$ is
present and the fact that the sewing ket is not changed.
Since the consistency of string field theory
depends only on the proper  BRST action on correlators and sewing,
the above arguments
indicate that there is no difficulty in defining string field theory
on the extended complex.

\section{Cohomology of $\Q$}

While the extended BRST complex is larger than the
original one, a priori,
this  has no immediate implication for the cohomologies. When we
extend a complex we increase both the number
of BRST closed states and the number of BRST trivial states.
As we saw earlier, some zero momentum states
that were physical in the original complex are trivial in the
extended one. On the other hand there
might be new solutions to $Q\ket\Psi=0$ in the new complex.  The
cohomology of the extended complex $\He$ is presented in
Ref.~[\astbel]. Indeed, we lose some states, in particular
states that do not change the physics of the background, and states
of peculiar ghost numbers. We gain
some states, but the new states can be understood in terms of the old
complex.

When we have a physical state $\ket{v,p}$ which
remains physical under  continuous variations of the momentum $p$, we
can easily construct physical states in the extended complex by taking
linear combinations of $\ket{v,p_i}$ where all $p_i \approx p$ are on-shell.
Adjusting the coefficients
we can  get as many derivatives with respect to $p$ as we want which
we can interpret as factors of $X_0^\mu$. Since mass-shells
are not flat, in general we  get nontrivial combinations of states
with different numbers of $X_0^\mu$.  At non-zero
momentum and ghost number two ($G=2$), all new physical states  can be obtained
from standard states by the above limiting procedure [\astbel] .

Let us  recall the structure of the semirelative
cohomology. At non-zero momentum $p$ the cohomology at
$G=2$ and at $G=3$  can be represented by the states
$c_1\ov c_1\ket{v,p}$, and $(c_0+\ov c_0)c_1\ov c_1\ket{v,p}$ respectively,
where $\ket{v,p}$ is a dimension $(1,1)$ primary matter state.
All $G=3$ states are trivial in the extended
complex. Indeed, using
Eq.~\qacts\ we can write
$$1\otimes (c_0+\ov c_0)c_1\ov c_1\ket{v,p}=
\Q\;{p\cdot x\over p^2}\otimes c_1\ov c_1\ket{v,p} -
\sum {p_\mu\over p^2}\otimes c_1\ov c_1(c_{-n}\a_n^\mu+\ov
c_{-n}\ov\a_n^\mu)\ket{v,p}\, .$$
The last sum must be $Q$ trivial because, being annihilated by $Q$,
$b_0$ and $\ov b_0$,  it would represent a
non-trivial relative cohomology class at $G=3$. Such a class does
not exist.

Calculation of the cohomology of the extended complex at zero momentum
is  more delicate [\astbel]. In the standard semirelative case the
physical states go as follows. At $G=2$
we have the $(d^2+1)$
states
of section three,  and at $G=3$
the $(d^2+1)$  states
obtained by multiplying the $G=2$ states by
$(c_0+\ov c_0)$. There are  $2d$ states at $G=1$ and
at $G=4$, and one state at $G=0$ ($SL(2,\IC)$ vacuum) and
at $G=5$.
In the extended complex there are no zero-momentum
physical states at $G>2$. There is an infinite
tower of $G=2$ states which can be described as different
limits  of linear combination of massless states as all momenta are
taken to zero. There is only one $G=0$ state, the
$SL(2,\IC)$ vacuum, and there are $d(d+1)/2$
physical states at $G=1$ which contain no more than one $X_0^\mu$. These
are precisely the states that generate Poincare symmetry. While
in the standard complex one gets the states that generate translations,
the states generating Lorentz transformations are missing. They appear
properly in the extended complex.

\chapter{CFT Deformations and the Matter Dilaton}

In this section we give a detailed analysis of the operator
$\partial X\bar\partial X$ and its effect on conformal theories
that include a free field $X$ living on the open line.
There are two cases of interest. In the first one,
the conformal theory includes the ghost system and has total
central charge of zero. We derive identities that show that
$\partial X\bar\partial X$
induces a trivial deformation of the CFT,
the ghosts playing a crucial role here.

We then consider the second case,
when this operator appears
in the context of the $c=1$  matter conformal field theory of the free
field (no extra ghosts). We use the definition of a
$c\not= 0$ conformal theory in the
operator formalism to show that the deformation {\it is not\/} strictly
trivial.
The detailed analysis shows  that the deformation in question can be
mostly eliminated by a change of basis in the conformal theory, but
the scale of the world sheet metric
is changed by the deformation.

The above results are certainly not controversial for the case of zero
central charge. The triviality of the  operator in question
has been argued earlier at various levels of detail. In
ref.[\mahapatra], for example, the usual argument that such perturbation
can be redefined away from the conformal field theory lagrangian is
reviewed, along with a discussion from the viewpoint of
gauge transformations in string field theory. In ref.[\mende]
the deformation of the two-point function of stress-tensors is investigated
explicitly. For the case of non-zero central charge our result appears
to be new.

We then turn to  integrals of
string forms over moduli spaces of Riemann surfaces
and consider their deformation by the insertion of the matter operator
in question.  The resulting integral expressions will be
needed in section six in order to establish the complete dilaton theorems.

\section{The operator $\partial X\bar\partial X$ in $c=0$ CFT}

Here we answer the following question: does the
$(1,1)$ primary field
$\np{\p X^\mu\bar\p X^\nu}$ define a
non-trivial deformation of a conformal field theory ?
Being a $(1,1)$ primary we can write the corresponding deformation
by integrating the field over the surface minus unit disks [\campbell,
\ranganathan]
$$ \delta^{\mu\nu}\, \bra{\Sigma_{g,n}}\, \equiv
\,{1\over 2\pi i} \int_{\Sigma_{g,n}-\cup D_k} \braket{\Sigma_{g,n+1}(z,\ov z)}
{\p X^\mu\bar\p X^\nu}\,\d z\wedge\d\ov z\,\, ,\eqn\cftdeff$$
This deformation is trivial if there is
an operator $\O^{\mu\nu}$ such that
$$\delta^{\mu\nu}\,
\bra{\Sigma_{g,n}} = -\bra{\Sigma_{g,n}\, }\sum_{k=1}^n\O^{\mu\nu\,(k)} ,
\eqn\expectDx$$
since this means that the deformation can be absorbed by a change of basis
in the CFT, a change induced by the operator $\O^{\mu\nu}$.
We will show that $\O^{\mu\nu}$ is given by
$$\O^{\mu\nu}= \dx{\mu\nu}+
     \, {1\over 6}\,
\displaystyle\met^{\mu\nu}\,G\,,\eqn\whytriv$$
where $\dx{\mu\nu}$ was defined in \dxints\ and $G$ is the total
ghost number operator.
This shows that the the operator
$\np{\p X^\mu\bar\p X^\nu}$ induces a trivial conformal field theory
deformation.

We now give a simple proof of the above assertion. Note that the
operator-valued two form that is being integrated can be written as
$$ \np{\p X^\mu\ov\p X^\nu}\d z\wedge\d \bar z =
   \d\,\Big[\,\half\, \np{ X^\mu\bar \p X^\nu}
\d \bar z-\half \np{ X^\nu\p X^\mu} \d z
       + \big(A^{\mu\nu}(z)\d z - \ov A^{\mu\nu}(\ov z)\d\ov z\big)\Big]\,,
\eqn\attep$$
where $A^{\mu\nu}(z)$ and $\ov A^{\mu\nu}(\bar z)$ are  arbitrary holomorphic
or antiholomorphic operators, and are therefore are annihilated by the exterior
derivative. They are important, however. The left hand side of the above
equation is a well defined two-form, but the expression within brackets
in the right hand side is not a well-defined one-form unless the $A$ operators
are suitably chosen. This is the case
because the operators  $\np{X^\mu\bar \p X^\nu}$
and $\np{ X^\nu\p X^\mu}$ are not primary. We can obtain primary operators
by choosing non-primary $A$ and $\ov A$ operators. We take
$$\np{\p X^\mu\ov\p X^\nu}\d z\wedge\d \bar z =
   \d\, \Big[\half\,\np{ X^\mu\bar \p X^\nu}\d \bar z
-\half \np{ X^\nu\p X^\mu}\d z
       + {\met^{\mu\nu}\over6}\big(G(z)\d z - \ov G(\ov z)\d\ov z\big)\Big]\,,
\eqn\nowr$$
where
$G(z) =\np {c(z)b(z)}$ and $\ov G( \ov z) =\np {\ov c(\ov z) \ov b(\ov z)}$
are the holomorphic and antiholomorphic ghost currents.
Indeed, since $T(z)G(w)\sim -3/(z-w)^3 +\cdots$,
the anomaly in the transformation law of
$:\hskip-3pt X^\nu\p X^\mu\hskip-3pt :$ is
canceled (see Eq.~\trans).
The expression inside the parenthesis is now a well-defined one form.
This allows us to use Stokes's theorem to convert the integral over the
surface minus the disks to an integral over the disks. Therefore, the
original expression \cftdeff, written in correlator language as
$${1\over 2\pi i}\int_{\Sigma_{g,n}-\cup D_k} \Bigl\langle\, \cdots
\np{\p X^\mu \ov \p X^\nu} \d z\wedge \d \ov z \,\,\cdots \Bigr\rangle\,,
\eqn\orig$$
becomes
$${1\over 2\pi i}
\sum_{k=1}^n \Bigl\langle \,\,\cdots\,\,\oint_{\p D_k}
\half\big( \np{ X^\nu\p X^\mu}\d z-\np{X^\mu\ov\p X^\nu}\d\ov z\big)^{(k)}
    - {\met^{\mu\nu}\over6}\big(G(z)\d z
      - \ov G(\ov z)\d\ov z\big)^{(k)}\,\,\cdots \Bigr\rangle
\eqn\primrep$$
where the contour integrals are over the boundaries of the disks oriented
as such.
We now recognize that the contour integrals
simply represent a single operator acting on each puncture, one at a time.
The operator is just
$${1\over 2\pi i}\oint_{|z|=1}
\half\big(\np{ X^\nu\p X^\mu}\d z-\np{ X^\mu\bar \p X^\nu}\d\ov z\big)
-{\met^{\mu\nu}\over6}\big(G(z)\d z
-\ov G(\ov z)\d\ov z\big) = -\Bigl( \dx{\mu\nu}+\met^{\mu\nu}{G\over 6}
\Bigr) \, .
\eqn\primrep$$
This concludes our proof of the triviality of the deformation.

\section{The operator $\partial X\bar\partial X$ in $c\not= 0$ CFT}

If we have any matter conformal theory coupled
to the ghost conformal theory, the ghost number operator $G$ acting on
surface states will give
$$\bra{\Sigma_{g,n}}\sum_{k=1}^nG^{(k)} = 6(1-g)\bra{\Sigma_{g,n}}.
\eqn\ghostano$$
Using this result we recognize that the result of the previous subsection
showing the triviality of the deformation can be written as
$$\eqalign{
\delta \, \bra{\Sigma_{g,n}} &=\,{1\over 2\pi i}
\hskip-10pt\int_{\Sigma_{g,n}-\cup D_k}
\hskip-10pt\braket{\Sigma_{g,n+1}(z,\ov z)}
{\p X^\mu\bar\p X^\nu}\,\d z\wedge\d\ov z\,\, \cr
&= -\bra{\Sigma_{g,n}\, }\sum_{k=1}^n  \dx{\mu\nu}^{(k)} -
\, (1-g)\,\met^{\mu\nu}\,\bra{\Sigma_{g,n}\,}\,\,,\cr} \eqn\whynot$$
In this form, of course, the triviality of the deformation
is not manifest since the second term in the right hand side is
not written as a sum of linear operators acting on the
surface state.

We now claim that equation \whynot\  applies
for the case when the matter conformal theory does not
include the ghost conformal theory. By construction, \whynot\
applies when the total conformal theory is the $c=1$ matter
theory coupled to the ghosts. The surface states in this total conformal
theory are the tensor product of the surface states in the two separate
conformal theories
$\bra{\Sigma_{g,n}\, }= \bra{\Sigma_{g,n}^{c=1}\, }\otimes\bra{\Sigma_{g,n}
^{gh}\, }$. Since the operators in the right hand side of \whynot\
are ghost-independent it is clear that we can factor out the ghost part
$\bra{\Sigma_{g,n}^{gh}\, }$ of the surface state.
Since the insertion in the left hand side carries no ghost dependence
the additional puncture is deleted in the ghost sector
$\braket{\Sigma_{g,n+1}(z,\ov z)}
{\p X^\mu\bar\p X^\nu}\,= \braket{\Sigma_{g,n+1}^{~c=1}(z,\ov z)}
{\p X^\mu\bar\p X^\nu}\otimes \bra{\Sigma_{g,n}^{gh}\, }$. It follows that
we can factor the ghost sector out of equation \whynot\ totally,  and
we find
$$\eqalign{
\delta \, \bra{\Sigma_{g,n}^{c=1}\,} &=\,{1\over 2\pi i}
\hskip-10pt\int_{\Sigma_{g,n}-\cup D_k}
\hskip-10pt\braket{\Sigma_{g,n+1}^{~c=1}(z,\ov z)}
{\p X^\mu\bar\p X^\nu}\,\d z\wedge\d\ov z\,\, \cr
&= - \bra{\Sigma_{g,n}^{c=1}\, }\sum_{k=1}^n  \dx{\mu\nu}^{(k)} -
     \, (1-g)\,
\met^{\mu\nu}\,\bra{\Sigma_{g,n}^{c=1}\,}\,\,.\cr} \eqn\whycot$$

We can now address the issue of triviality. As mentioned earlier, the
first term in the right hand side is just a similarity transformation.
The second term is not. To give an interpretation to that term we recall
the scaling properties of $c\not= 0 $ CFT. Under a scale change of
the metric on the surface
the correlators change as
$$\Bigl\langle  \cdots \Bigr\rangle_{\hbox{g} e^\sigma} =
\exp\, \Bigl[ {c\over 48\pi}
S_L (\sigma ; \hbox{g})\Bigr] \,
 \Bigl\langle  \cdots \Bigr\rangle_{\hbox{g}} \eqn\howscale$$
where
$$S_L (\sigma ; \hbox{g}) = \int d^2\xi \sqrt{\hbox{g}}\,
\Big(\half {\hbox{g}}^{\alpha\beta}
\partial_\alpha \sigma \partial_\beta \sigma  + R(\hbox{g})\sigma\Bigr)
\eqn\likethis$$
For constant $\sigma$ we get $S_L (\sigma ; \hbox{g}) =\sigma \int d^2\xi
\sqrt{\hbox{g}}\,
 R(\hbox{g}) =  4\pi \sigma\, (1-g) $, where $g$ is the genus of the
surface. This shows that for an infinitesimal scaling parameter
$\sigma$  the correlators scale as
$$\Bigl\langle  \cdots \Bigr\rangle_{\hbox{g} e^\sigma} =
\Bigl(1+ {c\over 12} \,\sigma (1-g) \Bigr) \,
\Bigl\langle  \cdots \Bigr\rangle_{\hbox{g}}\,. \eqn\scaletr$$
This shows that the last term in \whycot\ corresponds to a constant
scale deformation of the metric on the two-dimensional surface. The
deformation induced by $\partial X\bar\partial X$ is not completely trivial.

\section{Generalization to spaces of surfaces}

We must now extend the  discussion of the $c=0$ case
to include spaces of surfaces. Since we will use the ghost sector
in a nontrivial fashion we use the ghost number two
primary states $D^{\mu\nu}$ defined in \Dmn.
Rather than integrating the matter dilaton
over a single surface $\Sigma$, an operation that we can denote as
$f_{D^{\mu\nu}} (\K\Sigma)$, we want to consider the object
$f_{D^{\mu\nu}} (\K\A)$, where $\A$ is a space of surfaces.
We claim that the following result holds
$$f_{D^{\mu\nu}} (\K\A) + f_{\chi^{\mu\nu}}(\L \A)
      = \half\, \met^{\mu\nu}\, (2g-2+n)f(\A) \,,
\eqn\Dmth$$
where $\chi^{\mu\nu}$ is the state defined in \ximn\ and whose
main property is that upon action by the BRST operator it gives us
the matter dilaton state. The purpose of the present subsection is
to prove equation \Dmth.

We begin the proof of the above relation
by evaluating $f_{D^{\mu\nu}}(\K \A)$. Since the matter dilaton state can
be written as
$\ket{D^{\mu\nu}} = - Q \ket{\xi^{\mu\nu}} + \half \eta^{\mu\nu} \ket{D_g}$
we find
$$f_{D^{\mu\nu}}(\K \A) = -{1\over n!} \int_{\K\A} \bra{\Omega^{[\A + 2]g,n+1}}
Q \ket {\xi^{\mu\nu}} + \half \,\eta^{\mu\nu} f_{D_g} (\K\A)\,.\eqn\gensurf$$
Using the relation $f_{D_g}(\K\A) + f_{\chi_g} (\L \A) = (2g-2-n) f(\A)$
[\bergmanzwiebach],
we rewrite the above as
$$f_{D^{\mu\nu}}(\K \A)
+ \,\half\, \eta^{\mu\nu} f_{\chi_g} (\L\A)
 = -{1\over n!} \int_{\K\A} \bra{\Omega^{[\A + 2]g,n+1}}
Q \ket {\xi^{\mu\nu}} + \,\half \,\eta^{\mu\nu} (2g-2+n)f(\A) \,.\eqn\gsf$$
By virtue of \ximn, we see that \gsf\ implies the desired result \Dmth\ if
$$f_{\xi^{\mu\nu}}(\L\A) =  {1\over n!}
\int_{\K\A} \bra{\Omega^{[\A + 2]g,n+1}}
Q \ket {\xi^{\mu\nu}} \,.\eqn\nowprv$$
We must therefore establish this equation. Our first step is to replace the
BRST operator, which is only acting on the additional puncture,
by a sum of BRST operators acting on all punctures. The right hand side
then becomes
$${1\over n!} \int_{\K\A} \bra{\Omega^{[\A + 2]g,n+1}}
\sum_{k=1}^{n+1} Q^{(k)} \ket {\xi^{\mu\nu}}-
\, {1\over n!} \int_{\K\A} \bra{\Omega^{[\A + 2]g,n+1}}
\xi^{\mu\nu}\rangle \sum_{k=1}^n Q^{(k)}  \,. $$
The  second integral can be seen to vanish identically.
It involves the motion of the insertion over fixed
surfaces, and thus includes two antighost
insertions that have the property of annihilating
the vacuum state.  Since the $\xi^{\mu\nu}$ state only has a single ghost
operator acting on the vacuum, the two antighost insertions will annihilate it.
The first integral is rewritten by using the BRST property
of forms and Stokes's theorem
$$ {1\over n!} \int_{\K\A} \d\, \bra{\Omega^{[\A + 1]g,n+1}}
\xi^{\mu\nu}\rangle  = {1\over n!} \int_{\p (\K\A)} \bra{\Omega^{[\A +
1]g,n+1}}
\xi^{\mu\nu}\rangle \,. \eqn\givena$$
We now recall that
$\p(\K\A) = \K (\p\A) + \L \A$. The integral over $\K (\p\A)$ vanishes for
exactly the same reason as quoted in the above paragraph; two antighosts
insertions for position that annihilate the state. Thus the above term is
simply
$${1\over n!} \int_{\L\A} \bra{\Omega^{[\A + 1]g,n+1}}
\xi^{\mu\nu}\rangle = f_{\xi^{\mu\nu}} (\L\A)\,. $$
This establishes the correctness of \nowprv, and as a consequence
concludes our proof of \Dmth.

\chapter{Complete Dilaton Theorem}

In this section we  write a general hamiltonian
that induces string field diffeomorphisms relevant for the
dilaton theorem. Such hamiltonian will take a form similar to
that of Ref.[\bergmanzwiebach] and will allow us to treat in
a uniform way the dilaton, the ghost-dilaton, the matter-dilaton
and the graviton-trace states. We explain what kind of deformations
these various states produce, and emphasize that none of them leads
to a change of the slope parameter $\alpha'$. We then establish the complete
dilaton theorem for critical closed string field theory. The main point is
that the complete dilaton state can be written as
$$\ket{D} = - Q \,\ket{\,\xi^\mu_\mu}\, + {(d-2)\over 2}\ket{D_g}\,,\eqn\ytu$$
and therefore in the cohomology of the extended complex
the complete dilaton is just proportional to the ghost-dilaton
$$\ket{D} \approx   {(d-2)\over 2} \ket{D_g}\,. \eqn\oopss$$
In the extended complex only the ghost-dilaton changes the string
coupling, and the above equation indicates that the complete dilaton
changes the string coupling with a proportionality factor $(d-2)$.

\section{A General Hamiltonian}

What we have in mind here is writing a hamiltonian ${\bf U}$
that generates a diffeomorphism $F$ of the string field
via a canonical transformation
$$F :  \ket{\Psi} \,\to\, \ket{\Psi}
 + dt\,\{\, \ket{\Psi} ,{\bf U}\, \} \, . \eqn\infi$$
Associated to such canonical
transformation we can imagine that a parameter $\lambda$ of the string
measure is shifted. This is expressed as
$$ F^* \left\{ d\mu (\lambda) \exp \left({ 2\over \hbar} S (\lambda) \right)
\right\} \, =d\mu(\lambda + d\lambda) \exp \left({ 2\over \hbar}
S (\lambda + d\lambda ) \right)
\, ,\eqn\nud$$
namely, the diffeomorphism pulls back the relevant
{\em measure} of the theory with parameter $\lambda$ to
the measure of the theory with parameter $\lambda + d\lambda$. In order to
express the requirement \nud\ explicitly we use the following two relations
[\senzwiebachtwo]
$$\eqalign{
F^* (d\mu(\lambda)) &= {\rho(\lambda)\over \rho(\lambda + d\lambda)}\,
d\mu(\lambda + d\lambda )
\,\left( 1 + 2\,dt\,\Delta{\bf U}\right) \, ,\cr
F^* \{ S(\lambda)\} &= S(\lambda) + dt\, \{ S( \lambda) ,
{\bf U} \}\,,\cr}\eqn\tractd$$
where $d\mu (\lambda) = \rho(\lambda) \prod d\psi$.
Equation \nud\ then reduces to
$$\Bigl( {d\lambda\over dt}\Bigr) \cdot \, {d\over d\lambda}
\left( S + {1\over 2} \hbar \ln \rho\right)
 =  \{ S , {\bf U}\, \} +
\hbar \Delta {\bf U} \equiv \hbar
\Delta_S {\bf U}\, .\eqn\mainc$$
If the right hand side of the above equation is zero for
some specific hamiltonian ${\bf U}$, we must conclude that
the diffeomorphism does not change anything in the string field measure.
The diffeomorphism is then  a symmetry transformation.

The  hamiltonian we will introduce depends on
a pair of states $\O$ and $\chi$ related by a BRST operator:
$$\O+\{\QQ,\chi\}=0\,  \quad\to\quad \quad  \ket{\O} = - Q \ket{\chi} .
\eqn\rel$$
We demand that  $\ket\O\in \H$, namely
$b_0^-\ket{\O}= L_0^- \ket{\O} =0$, but $\ket\chi$
need not be annihilated by $b_0^-$, and
may involve the coordinate operator.
Neither state needs to be primary.
We define
$${\bf U}_{\O,\chi} =  {\bf U}_\O^{[0,2]} - f_\O(\B_>)
                   + f_\chi(\V_{0,3} +\{\B_{0,3},\V\}).\eqn\UOeta$$
Note that this hamiltonian may involve a state $\chi$ which is {\it not}
annihilated by $b_0^-$ since this state only appears inserted on three
punctured spheres and there is no problem defining the phase of a local
coordinate on such collection of surfaces (see [\bergmanzwiebach]).
Using this hamiltonian we will be able to treat ghost and matter
dilatons in a similar fashion.
We now follow  [\bergmanzwiebach] to compute the right hand side of \mainc.
Since the computation is rather similar, we will be brief.
The first term of the right hand side gives
$$
\eqalign{
 \{ S , {\bf U}_{\O,\chi} \} =
    & \quad {1\over \k} \,\{ S , {\bf U}^{[0,2]}_{\O} \}
    +\{ S , f_\chi ( \underline\V_{0,3}) \} \cr
    & -{1\over \k} \, \{ S ,  f_\O ( \B_> )\}
    + \{ \, S ,\;  f_\chi ( \{ \B_{0,3} , \V \})\,\, \}\,. \cr}
\eqn\asic
$$
Making use of the identities given in Eqs. (2.47) and (2.48) of
Ref.[\bergmanzwiebach] we can rewrite \asic\ as
$$
\eqalign{
 \{ S , {\bf U}_{\O,\chi} \} =
    & \,\,\,\,{1\over \k} \,f_\O ( \underline\V )
    - f_\O ( \underline\V_{0,3})
         - f_\chi ( \{ \V , \underline \V_{0,3} \} )
    + {1\over \k} \,  f_\O \Bigl(\partial\B_>
         + \{ \V , \B_>\} \Bigr) \cr
    &- f_\O (\{ \B_{0,3} , \V \}) -
         f_\chi\Bigl( \partial \{ \B_{0,3} , \V \}
           + \{ \V , \{ \B_{0,3}\,, \V \}    \} \Bigr) \,, \cr}
\eqn\assic
$$
where each row in \assic\ is equal to the corresponding row in \asic.
Using the Jacobi identity
in the fourth row, the geometrical recursion relations,
and the definition $\p\B_{0,3} = \V'_{0,3}- \underline\V_{0,3}$, we obtain
$$\{ S , {\bf U}_{\O,\chi} \} = {1\over \k} \,f_\O \Bigl(
 \partial\B_>  + \{ \V , \B \} + \underline\V_> \Bigr)
+  f_\chi \Bigl( \{ \B_{0,3} , \hbar\Delta\V\}
            -\{\V'_{0,3},\V\}\Bigr)\, .\eqn\assicx$$
A similar calculation gives
$$ \hbar\Delta {\bf U}_{\O,\chi}
        = -\,\hbar f_\chi\, (\Delta \underline\V_{0,3}) + {1\over \k}\, f_\O
(\hbar\Delta\B_>) -\, f_\chi\,(\{ \B_{0,3}\,, \hbar \Delta \V \} )\,.
\eqn\nbv$$
Equations \nbv\ and \assicx\ must be added to give the right hand side
of \mainc. Doing this, and using the recursion relations
$$\p\B_>\, =
    \k\ov\K\V- \hbar\Delta\B_> - \{\V,\B\} - \underline\V_>
     + \hbar\k\,\underline\V_{1,1}\, ,\eqn\improvett$$
for the $\B$ spaces we finally find
$$\fbox{$\ds
     \Delta_S \,{\bf U}_{\O,\chi} =  \,f_\O (\ov\K\V) + f_\chi(\ov \L\V) +
     \hbar\, [ \, f_\O (\underline\V_{1,1})
     -\,f_\chi(\Delta \underline\V_{0,3})\, ].$}\eqn\aseeic$$
Note that by definition,  the term $\ov\L\V$ does not include
vertices with zero punctures. Writing out the above equation more explicitly
we have
$$\eqalign{
     \Delta_S \,{\bf U}_{\O,\chi} &=  \sum_{g, n\geq 1}
\,f_\O (\ov\K\V_{g,n}) + f_\chi(\ov \L\V_{g,n})\cr
& + \sum_{g\geq 2}
\,f_\O (\ov\K\V_{g,0})  \cr
&+    \hbar\, [ \, f_\O (\underline\V_{1,1})
     -\,f_\chi(\Delta \underline\V_{0,3})\, ].\cr}\eqn\aexok$$

\section{Application to the various dilaton-like states}

In this section we will use the general hamiltonian \UOeta\ and its
basic property \aseeic\ to calculate the effect of shifts of
the ghost dilaton $\Dg$, the matter dilaton $D^\mu_\mu$,
the true dilaton $D$, and
the graviton trace ${\cal G}_t$.
We begin with the case of the ghost-dilaton, fully
analyzed in Refs.[\bergmanzwiebach,\rahmanzwiebach], as a way to use
the present general formalism.

\noindent
\underbar{Ghost-Dilaton.}~Since
$\Dg+\{\QQ,\Xg\} = 0$, we consider the hamiltonian
${\bf U}_{D_g,\chi_g}$ (a simpler form of the ghost-dilaton hamiltonian
will be given in the next section). The identities
$$f_\Dg(\K\V_{g,n})+f_\Xg(\L\V_{g,n})=(2g-2+n)f(\V_{g,n}),\eqn\dgth$$
established in Ref.[\bergmanzwiebach], and the identities
$$\eqalign{
f_\Dg(\K\V_{g,0})& =(2g-2)f(\V_{g,0})\,,\quad g\geq 2\,,\cr
 f_{\D_g} (\underline\V_{1,1}) &=
     f_{\chi_g} (\Delta \underline\V_{0,3}) = 0\,, \cr}\eqn\srah$$
established in [\rahmanzwiebach] imply that \aexok\ yields
$$\Delta_S \,{\bf U}_{\D_g,\chi_g} =  \sum_{g, n}\hbar^g \kappa^{2g-2+n}
\,(2g-2+n) f(\V_{g,n})\,.
\eqn\aek$$
Since the string action is given by
$S= S_{0,2} + f(\V) + \hbar S_{1,0}$
and the kinetic term $S_{0,2}$, the elementary vacuum term $S_{1,0}$,
and the measure $\ln \rho$ are all coupling constant independent
we can write \aek\ as
$$\Delta_S \,{\bf U}_{\D_g,\chi_g} = \kappa\, {d\over d\kappa}\,
 \left( S + {1\over 2} \hbar \ln \rho\right)\,.
\eqn\aekk$$
This equation shows that the ghost dilaton changes the coupling constant
$\kappa$. Comparing with \mainc\ we see that ${d\kappa\over dt}=\kappa$,
or $\kappa = \kappa_0 e^t$. Here $t$ plays the role of the vacuum expectation
value of the ghost-dilaton.

\noindent
\underbar{Matter dilaton.}~ Since $D^{\mu\nu} +\{Q,\chi^{\mu\nu}\}  = 0$,
we are led to consider the
hamiltonian ${\bf U}_{\D^{\mu\nu},\chi^{\mu\nu}}$. This time \Dmth\ is
relevant in the form
$$f_{D^{\mu\nu}} (\K\V_{g,n}) + f_{\chi^{\mu\nu}}(\L \V_{g,n})
      = \half\, \met^{\mu\nu}\, (2g-2+n)f(\V_{g,n}) \,   .
\eqn\qmth$$
Moreover we claim that
$$f_{D^{\mu\nu}} (\K\V_{g,0})
      = \half\, \met^{\mu\nu}\, (2g-2)f(\V_{g,0}) \,  \quad g\geq 2 .
\eqn\rmth$$
This follows from
$\ket{D^{\mu\nu}} = - Q \ket{\xi^{\mu\nu}} + \half \eta^{\mu\nu} \ket{D_g}$,
the first equation in  \srah, and the vanishing of
$f_{Q\xi^{\mu\nu}} (\K\V_{g,0})$.
On the other hand
$$ f_{{\cal G}^{\mu\nu}} (\underline\V_{1,1}) -\,f_{\xi^{\mu\nu}}
 ( \Delta \underline\V_{0,3}) = 0
\,\,,\eqn\cex$$
by virtue of \gravdef\ and Stokes theorem which is valid here because
$b_0^-$ annihilates the state $\xi^{\mu\nu}$. Since
$D^{\mu\nu} = {\cal G}^{\mu\nu} + \half \eta^{\mu\nu} D_g$,
it now follows that
$$\eqalign{
f_{D^{\mu\nu}}(\V_{1,1}) &= f_{{\cal G}^{\mu\nu}}(\V_{1,1}) +
\half\eta^{\mu\nu} f_{D_g}(\V_{1,1})\,,\cr
&= f_{\xi^{\mu\nu}}
 ( \Delta \underline\V_{0,3})\,\,,  \,
\quad\qquad\qquad (\hbox{Using}\, \srah\,\,
\hbox{and}\,\,\cex )   \cr
&= f_{\chi^{\mu\nu}} ( \Delta \underline\V_{0,3}) -{1\over 2}\eta^{\mu\nu}
f_{\chi_g} (\Delta \underline\V_{0,3})\,, \qquad (\hbox{Using}\,
\ximn) \cr
&=  f_{\chi^{\mu\nu}} ( \Delta \underline\V_{0,3})\,\,,
\qquad\qquad\qquad\qquad\qquad (\hbox{Using}\,\srah)\, \cr} \eqn\extd$$
and therefore
$$f_{D^{\mu\nu}}(\V_{1,1})-f_{\chi^{\mu\nu}}
( \Delta \underline\V_{0,3})=0\,.\eqn\getr$$
The computation of $\Delta_S {\bf U}_{D^{\mu\nu},\chi^{\mu\nu}}$
is now straightforward. The terms in the right hand
side of \aexok\  have been evaluated in Eqs.\qmth,\rmth, and \getr.
We then find
$$\Delta_S \,{\bf U}_{D^{\mu\nu},\chi^{\mu\nu}} =
 \half \eta^{\mu\nu}\,\cdot \kappa\, {d\over d\kappa}\,
 \left( S + {1\over 2} \hbar \ln \rho\right)\,.
\eqn\aeykk$$
Note that the off-diagonal states $(\mu \not= \nu)$ have no effect whatsoever,
they ought to be interpreted as generating gauge transformations.
Each one of the $d$ diagonal states change the coupling constant.
In particular, for the trace state we have
$$\Delta_S \,{\bf U}_{D^\mu_\mu,\chi^\mu_\mu} =
 {d\over 2}\,\cdot \kappa\, {d\over d\kappa}\,
 \left( S + {1\over 2} \hbar \ln \rho\right)\,.
\eqn\aeykkp$$
The only effect of a shift of the matter dilaton
$D^\mu_\mu$ is a shift of the string coupling, with a strength
proportional to the number of noncompact dimensions.

\noindent
\underbar{The complete dilaton.}~ This state
is written as
$D = D_\mu^\mu  -D_g = - \,\{ Q\, , \, \chi^\mu_\mu - \chi_g \}$.
Therefore
$$\Delta_S {\bf U}_{D, \chi_d} = \Delta_S {\bf U}_{D_\mu^\mu, \chi_\mu^\mu}
- \Delta_S {\bf U}_{D_g, \chi_g} = \Bigl( {d-2\over 2}\Bigr)
\cdot \kappa\, {d\over d\kappa}\,
 \left( S + {1\over 2} \hbar \ln \rho\right)\,,\eqn\cdild$$
where use was made of \aeykkp\ and \aekk. This equation shows that the only
effect of a  dilaton shift is changing the string coupling with
a strength proportional to the number of noncompact dimensions minus two.
This is the complete dilaton theorem in critical bosonic closed string
field theory.

It is sometimes said that the effect of a dilaton is to change the
dimensionless string coupling {\it and} the slope parameter $\alpha'$.
We believe such statements are at best misleading. We see in the
above discussion that  $\Delta_S {\bf U}$ amounts to just
changing the dimensionless string coupling. The slope parameter is
the only dimensionful parameter in the string theory, and, as such, it is
not really a parameter but a choice of units. There is
no invariant meaning to a change in $\alpha'$. If the theory had
another dimensionful parameter, say a compactification radius $R$, then
there is a new dimensionless ratio $R/\sqrt{\alpha'}$ that can be changed
in the theory. One can view such change, if so desired, as a change of
the dimensionful radius of compactification, or equivalently, as a change
in $\alpha'$. Still, it should be remembered that only changes
in dimensionless couplings have invariant meaning.

\noindent
\underbar{Graviton Trace.}~
The final case of interest is that of the states ${\cal G}^{\mu\nu}$ written
as
$${\cal G}^{\mu\nu} = D^{\mu\nu} - \half \eta^{\mu\nu} D_g = - \,\{ Q\,,
\, \chi^{\mu\nu} - \half\eta^{\mu\nu} \chi_g \, \}\,. \eqn\thgio$$
Therefore
$$\Delta_S {\bf U}_{{\cal G}^{\mu\nu}, \xi^{\mu\nu}}
= \Delta_S {\bf U}_{D^{\mu\nu}, \chi^{\mu\nu}}
- \half \eta^{\mu\nu} \Delta_S {\bf U}_{D_g, \chi_g} = 0\,,\eqn\cdilg$$
where use was made of Eqs.\aeykk\ and \aekk. This shows that none of
the  ${\cal G}^{\mu\nu}$ states deforms the string background. In
particular, the graviton trace ${\cal G}$
does not deform the string background. Note that the hamiltonian
$ {\bf U}_{{\cal G}^{\mu\nu}, \xi^{\mu\nu}}$ defines a string field
transformation whose inhomogeneous term is indeed ${\cal G}^{\mu\nu}$,
and leaves the string measure invariant. Our discussion of the extended
complex indicates that a completely equivalent field transformation
would be a gauge transformation generatred by $\xi^{\mu\nu}$.

\chapter{The Relevance of the Ghost-dilaton}

In this section we wish to consider the case when
the ghost-dilaton becomes a trivial state in the standard semirelative
BRST cohomology. This is not a hypothetical situation, it happens in
$D=2$ string theory, as will be reviewed in sect.8.
If the ghost dilaton is trivial it may seem that it cannot
change the coupling constant, leaving the possibility
that other states change it. This is not the way things work out. Inspection
of Ref.[\bergmanzwiebach] reveals that the ghost-dilaton shifts the
string coupling whether or not it is BRST trivial. It then seems clear
that the string coupling should not be an observable. We give a proof that
this is the case. Explicitly this means the following: while the string
coupling
is a parameter appearing in the string action its value can be changed
by a string field redefinition having no inhomogeneous term.

If the ghost dilaton is trivial, it can
be written as $\ket{D_g} = -Q\ket{\eta}$ with $\ket{\eta}$ a legal state
in the standard semirelative complex. If the two-form
associated to the motions of the state $\ket{\eta}$ vanishes, as is the case
for
$D=2$ strings, the field redefinition changing the string coupling
is simply a homogeneous field redefinition. This will be the
case whenever $\ket{\eta}$ involves a single ghost field acting on the
vacuum. If this is not the case,
the field redefinition is nonlinear; while it lacks an inhomogeneous term
at the classical level, there may be $\hbar$-dependent inhomogeneous terms.
We do not know of an example where $\ket{\eta}$ is this
complicated.  If this happens the string coupling might not be completely
unphysical at the quantum level.

\section{Simplifying the ghost-dilaton hamiltonian}
In this subsection we examine again the ghost-dilaton hamiltonian and
show that it can be simplified considerably. The simplified hamiltonian
will be of utility to show that a trivial ghost-dilaton implies an
unphysical coupling constant.

The ghost-dilaton hamiltonian reads
$${\bf U}_{D_g} =  {\bf U}_{D_g}^{[0,2]} - f_{D_g}(\B_>)
                   + f_{\chi_g} (\V_{0,3} +\{\B_{0,3},\V\}).\eqn\gdham$$
We now show that the last term in this hamiltonian, involving $\chi_g$
can be replaced by a simpler term involving the ghost-dilaton.
Consider the evaluation of $\Delta_S f_{\chi_g} (\B_{0,3})$
$$\eqalign{
\Delta_S f_{\chi_g} (\B_{0,3}) &= \Delta f_{\chi_g} (\B_{0,3})+ \{ S \,, \,
f_{\chi_g}(\B_{0,3})\} \,,\cr
&= -f_{\chi_g} (\Delta \B_{0,3}) -f_{Q\chi_g} ( \B_{0,3})
 -f_{\chi_g} (\partial \B_{0,3} + \{ \V \, ,\, \B_{0,3} \}\,) \,,\cr
&= -f_{\chi_g} (\Delta \B_{0,3}) +f_{D_g} ( \B_{0,3})
 +f_{\chi_g} ( \V_{0,3} + \{ \B_{0,3} \, ,\, \V \}\,) \,,\cr
} \eqn\replace$$
where we made use of the relation $f_{\chi_g} (\V'_{0,3})=0$, which follows
from sect.6.2 of Ref.[\bergmanzwiebach]. Rearranging the terms in
the equation we write
$$f_{\chi_g} ( \V_{0,3} + \{ \B_{0,3} \, ,\, \V \}\,) \,=\,
-f_{D_g} ( \B_{0,3})  \,+\, f_{\chi_g} (\Delta \B_{0,3})
\,+\, \Delta_S f_{\chi_g} (\B_{0,3}) \,.\eqn\rearr$$
Constant terms are irrelevant for hamiltonians since they do not
generate transformations. We can therefore drop the second term in the
above right hand side. Moreover, the third term in the right hand side
can also be dropped since it is annihilated by $\Delta_S$.
It follows that we can replace
$f_{\chi_g} ( \V_{0,3} + \{ \B_{0,3} \, ,\, \V \}\,)$ by
$(-f_{D_g} ( \B_{0,3}))$
in the dilaton hamiltonian. We thus find that
$${\bf U}_{D_g} =  {\bf U}_{D_g}^{[0,2]} - f_{D_g}(\B)
                   \,,\eqn\gdhamnew$$
is a hamiltonian equivalent to the original one,
and by a slight abuse of notation we denote it with the same symbol.
This hamiltonian is completely analogous to the background
independence hamiltonians found in Ref.[\senzwiebachtwo].
It is straightforward to verify that this hamiltonian has the desired
properties. One computes
$$\eqalign{  \Delta_S {\bf U}_{D_g} &=
\{ S , {\bf U}_{D_g}^{[0,2]} \} - \Delta f_{D_g} (\B )-
\{ S , f_{D_g} (\B) \}\,\,,\cr
&= f_{D_g} \Bigl( \partial \B + \{ \V \,, \, \B \} + \Delta \B \, + \,
\underline\V \Bigr) \,\,,\cr
&= f_{D_g} \Bigl( \,\overline \K \V + \V'_{0,3} + \Delta\B_{0,3} \,+ \,
\underline \V_{1,1}  \Bigr) \,\,,\cr
&=  f_{D_g} ( \overline \K \V  )\,,\cr}\eqn\verham$$
where use was made of the recursion relations \improvett\ together with
$\partial \B_{0,3} = \V'_{0,3} - \V_{0,3}$. In the last step we used
$f_{D_g} (\V'_{0,3}) =0$, which follows from $f_{\chi_g} (\V'_{0,3}) =0$
and the BRST property, and, of the result of [\rahmanzwiebach] that
the two-form associated to the ghost-dilaton vanishes
identically on the moduli space of once punctured tori. The fact that
the ghost-dilaton hamiltonian can be written in the standard background
independence form was not anticipated in [\bergmanzwiebach] because both
the identity $f_{D_g} (\V'_{0,3}) =0$, and the understanding of the
behavior of dilatons at genus one were missing.

\section{Triviality of the ghost-dilaton and the string coupling}

The ghost-dilaton theorem in string field theory, as established
in Refs.[\bergmanzwiebach,\rahmanzwiebach] holds true whether or
not the ghost-dilaton is BRST trivial or not. The
ghost-dilaton hamiltonian will always have the effect of changing the
string coupling. This ghost-hamiltonian produces a string field redefinition
that includes an inhomogeneous term, a shift along the ghost-dilaton
state. If the ghost-dilaton is trivial in the standard semirelative
BRST cohomology, then it can be written as $\ket{D_g} = - Q \ket{\eta}$,
where $\ket{\eta}$ is a standard vector in the closed string field theory
state space. It then follows that there is another hamiltonian, the
hamiltonian corresponding to a gauge transformation, that also has
the property of inducing a shift along the direction of the ghost-dilaton.
This gauge hamiltonian ${\bf U}_G$ reads
$$ {\bf U}_G = \Delta_S {\bf U}_\eta^{[0,2]} = {\bf U}_{D_g}^{[0,2]}
+ f_\eta(\V) \,, \eqn\ghamn$$
where the gauge invariance property follows from $\Delta_S {\bf U}_G=0$.
Since the gauge hamiltonian induces no change in the string action, it
follows that the hamiltonian
$${\bf U}_F \equiv
 {\bf U}_{D_g}- {\bf U}_G  = -f_{D_g} (\B) - f_\eta(\V)\,,\eqn\diffham$$
still shifts the string coupling constant.
The term ${\bf U}_{D_g}^{[0,2]}$ inducing the shift
along the ghost dilaton is absent in ${\bf U}_F$ and therefore the
hamiltonian ${\bf U}_F$ is a  hamiltonian that induces
a field redefinition without physical import; it does not change
the vacuum expectation value of the string field.
The fact that the string coupling
parameter in the action can be changed by a field redefinition without
an inhomogeneous term implies that the string coupling is unphysical.
As we will see now, strictly, and in all generality, this is only the
case for genus zero, or the $\hbar$-independent part of the string field
redefinition generated by ${\bf U}_F$. In order to appreciate this point
we now obtain a simple expression for the hamiltonian ${\bf U}_F$.

Our calculation begins by simplifying the expression for $f_{D_g}(\B)$ in
the ghost-dilaton hamiltonian by taking into account that
$\ket{D_g} = - Q \ket{\eta}$. We consider
$$\{ S \,,\, f_\eta (\B) \} = - f_{Q\eta} (\B) \,-\, f_\eta\,
\Bigl( \partial \B + \{ \V \,,\, \B \} \Bigr) \,. \eqn\simpham$$
Using $\Delta f_\eta (\B) = - f_\eta (\Delta \B)$, we rewrite the above
equation as
$$-f_{D_g} (\B) = - f_\eta\,
\Bigl( \partial \B + \{ \V \,,\, \B \} + \Delta \B \Bigr) -\Delta_S
\,f_\eta (\B) \,.\eqn\iuqd$$
Using the recursion relations \improvett\ we find
$$-f_{D_g} (\B) = - f_\eta\,
\Bigl( \V'_{0,3} + \overline\K \V - \underline\V \Bigr)
- f_\eta\,
\Bigl( \Delta \B_{0,3} + \V_{1,1} \Bigr) -\Delta_S
\,f_\eta (\B) \,.\eqn\iwfc$$
Since the left hand side is to be used in a hamiltonian we can drop the
second term, being a constant, and the last term, being annihilated by
$\Delta_S$. It follows that back in \diffham\ the hamiltonian
${\bf U}_F$ can be taken to be
$${\bf U}_F = -f_\eta\, ( \V'_{0,3} + \overline \K\V ) \,. \eqn\finans$$
This is the simplest form of the hamiltonian. We see that at genus zero
the hamiltonian is quadratic, and thus generates a field redefinition without
an inhomogeneous term. Thus, at genus zero it is completely clear that
the string coupling is unphysical.
At higher genus there are, in principle, non-vanishing
inhomogeneous terms arising from the surfaces $\overline\K \V_{g,1}$.
This might mean that the string coupling is not fully
unphysical at the quantum level. More likely, it may be that whenever
the ghost-dilaton is trivial the two-form corresponding to the motion of the
state $\ket{\eta}$ vanishes. This happens, for example, when each term
in $\ket{\eta}$ only has one ghost field acting on the vacuum, as
is the case in $D=2$ string theory. If the two-form vanishes
the contribution from $\overline\K \V$ vanishes as well, and the hamiltonian
${\bf U}_F  = -f_\eta\, ( \V'_{0,3} ) $ simply generates a
homogeneous field redefinition
$$\delta \,\ket{\Psi}_1 = -\{ f_\eta\, ( \V'_{0,3} ) \,,\,   \ket{\Psi}_1\}
= \Bigl[\, \bra{V'_{1'23}\,} \eta\rangle_3
\ket{{\cal S}_{11'}}\,\Bigr] \,\ket{\Psi}_2 \,.\eqn\fixs$$
The linear operator acting on the string field has the interpretation
of a contour integral of the conformal field operator $\eta(z,\bar z)$
using local coordinates induced by the special puncture in
the string vertex $\V'_{0,3}$.

It is a straightforward calculation using $ \ov \L = - \{ \V'_{0,3} \,, \V \} $
to show that acting on the first part of the hamiltonian \finans\ the operator
$\Delta_S$ gives
$$\eqalign{
\Delta_S \,\Bigl(\,  - \,f_\eta\, ( \V'_{0,3} ) \, \Bigr) &=
f_{D_g} (\V'_{0,3}) + f_\eta (\Delta\V'_{0,3} )  -
f_\eta (\ov\L \V)\,,\cr
&=  - f_\eta \,(\ov\L \V)\,. \cr} \eqn\ufff$$
In the last step we have used the vanishing of $f_\eta (\Delta\V'_{0,3} )$
which is readily established
$$\eqalign{
f_\eta \,(\Delta\V'_{0,3} ) &= f_\eta\, (\Delta\V_{0,3}- \partial
\Delta \B_{0,3} )\,\cr
&= f_\eta \, (-\partial \, [ \V_{1,1}- \Delta \B_{0,3}\,] ) \,\cr
&= f_{D_g} \, (\V_{1,1}- \Delta \B_{0,3})  = 0\,. \cr} \eqn\aside$$
Using once more the vanishing of the ghost-dilaton two-form on
the moduli space of punctured tori. It follows from \ufff\ and \finans\
that in general
$$\Delta_S \, {\bf U}_F =  - f_\eta \,(\ov\L \V) - \Delta_S
f_\eta\, (\overline \K\V )\,.\eqn\backcirc$$
Whenever the two-form associated with moving the state
$\ket{\eta}$ vanishes we see that
the effect of ${\bf U}_F$ reduces to inserting the state $\eta$ (via $\ov\L$)
on all the coordinate disks of the string vertices. One can readily verify
that for an arbitrary  $\ket{\eta}$ the relation
$\Delta_S \,f_\eta \,(\ov\K\V) = f_{Q\eta}\, (\ov\K\V) - f_\eta (\ov\L\V)$
holds. This confirms
that \backcirc\ is equivalent to $\Delta_S \, {\bf U}_F= f_{D_g} (\ov\K \V)$
as befits a hamiltonian that must change the string coupling.

\chapter{Application to  D=2 String Theory.}

In this section we consider the case of $D=2$ string theory as
an illustration of the ideas developed in the previous sections of
this paper. We analyze zero-momentum physical states that are candidates
for deformations of the string background. The semirelative BRST-physical
states are seen to be trivial in the extended complex and
should not deform the string
background.  We discuss why they do not appear to deform
the conformal theory. Our analysis here is a refinement of
that of Mahapatra, Mukherji and Sengupta [\mahapatra].  We argue that states
in the absolute cohomology that are not annihilated by $b_0^-$
do not lead to CFT deformations, thus evading a possible conflict with
background independence.  Finally, noting that
in this background the ghost-dilaton is
BRST trivial, we explain how to apply our earlier
considerations showing that the string coupling is unobservable.

\section{Zero-momentum states and CFT deformations }

Two dimensional string theory is
based on a matter CFT including  a Liouville field $\varphi (z,\bar z)$
and a field $X(z,\bar z)$. The
holomorphic matter energy-momentum tensor reads
$T_m = -\half\, \partial X \partial X -\half\partial\varphi
\partial \varphi + \sqrt 2 \,\partial^2 \varphi $.
Due to the background charge,
the field $\varphi$ does not
transform as a scalar. Under general analytic coordinate changes $z'(z)$
$$\varphi (z', \overline { z'})  = \varphi (z , \bar z) - 2\sqrt 2 \ln
\Bigl| {dz'\over dz}\Bigr| \,. \eqn\litr$$
In this string theory an important operator $a(z)$ [\li,\wittenzwiebach]
is obtained by taking the
commutator of the BRST operator with the field $\varphi$
$$a(z) \equiv {1\over \sqrt 2}\,
\{ Q\, ,\, \varphi (z, \bar z) \} =  \,\partial c+ {1\over \sqrt 2}
\, c\,\partial\varphi\, .\eqn\mliop$$
The operator $a(z)$ is  trivial  in semirelative cohomology of the
extended complex, but it is
nontrivial in the standard semirelative cohomology.

Let us consider the absolute cohomology BRST physical states
at ghost number two that can be formed
without using exponentials of the free fields.
The space of such states is spanned by a total
of six states [\wittenzwiebach], the first three of which are
states in the semirelative
cohomology
$$\eqalign{{\cal S}_1 &
= {1\over \sqrt 2}\,c\bar c \,\partial X \bar \partial\varphi +
c\partial X ( \partial c +\bar\partial \bar c) \,,  \cr
{\cal S}_2 & =
{1\over \sqrt 2}\,c\bar c \,\bar\partial X  \partial\varphi +
\bar c\bar \partial X ( \partial c +\bar\partial \bar c) \,, \cr
{\cal S}_3 & =c\bar c\partial X\bar\partial X\,,\cr}\eqn\dsrstate$$
and the other three are states that do not obey the semirelative condition
$$\eqalign{{\cal A}_1 & =
{1\over \sqrt 2}\,c\bar c \,\partial X \bar \partial\varphi +
c\partial X ( \bar\partial \bar c - \partial c) \,,  \cr
{\cal A}_2 & =
{1\over \sqrt 2}\,c\bar c \,\bar\partial X  \partial\varphi +
\bar c\bar \partial X ( \partial c -\bar\partial \bar c) \,, \cr
{\cal A}_3 & = a\,\cdot \bar a \,.\cr}
\eqn\dsrstates$$
The states in \dsrstate\ are the only states in the semirelative
cohomology at ghost number two
under the condition of zero momenta [\wittenzwiebach].

Let us begin our analysis with the first two
semirelative states. We first observe
that they are trivial in in the extended complex
$${\cal S}_1(z, \bar z)  =  \Bigl\{ Q\,  ,\, s_1  \Bigl\} \,,
\quad  s_1 = -\, {1\over \sqrt 2}\,\,
c \, \varphi \partial X \,,\eqn\taex$$
with an exactly analogous statement holding for ${\cal S}_2$. This indicates
 that such states do not deform the string
background. Indeed, since the BRST property holds for the class of states
containing $\varphi$, we may simply use the state $s_1$
as the gauge parameter in a string field gauge transformation.

As further confirmation that the string background
is not changed, let us now see that if we try to use the state ${\cal S}_1$ to
deform the underlying CFT the only possibility seems to
be  zero deformation. In order to use ${\cal S}_1$ to deform the CFT we must
find the corresponding coordinate invariant two-form.
We introduce a metric $\rho$ on the Riemann surface and a brief
computation using [\bergmanzwiebach] gives
$${\cal S}_1^{[2]}= -{1\over \sqrt 2} \, dz \wedge d\bar z\, \Bigl[
\partial X \, \bar \partial \,( \varphi - 2\sqrt 2 \ln \rho ) \,\Bigr] \, .
 \eqn\phis$$
We see that only the first term in ${\cal S}_1$
has contributed to the result.
Moreover, $\varphi$ has been replaced by the
coordinate invariant combination
$(\varphi - 2\sqrt 2 \ln \rho )$. While the two-form
is well-defined (it is coordinate invariant), it is not Weyl invariant.
The difficulty of obtaining a coordinate invariant and Weyl invariant
two-form was pointed out in [\wittenzwiebach]. At that time it was not clear
how to obtain the two-form associated to {\it arbitrary} states. We
now see that there is no Weyl-invariant two-form and thus
no well-defined way to integrate the two-form on Riemann surfaces.

The analysis cannot stop here. A similar
situation happens for the ghost-dilaton: its
two-form is not Weyl independent. This dependence drops out
of surface integrals when we add the integral of a
suitable one-form over the boundary of the region of integration.
That one-form is the one associated to the state $\ket{\chi_g}$
which acted by the
BRST operator gives the ghost-dilaton.
We  therefore consider the state $s_1$ defined in
\taex\ and construct the
corresponding one-form. We find
$$s_1^{[1]} = {1\over \sqrt 2} \, dz\,( \varphi -2\sqrt 2\ln\rho )
\partial X\, . \eqn\onef$$
This one-form is coordinate invariant  but it is
not Weyl invariant.
One readily verifies that ${\cal S}_1^{[2]} = d\, s_1^{[1]}$, as expected
from \taex. In the case of the ghost-dilaton the gauge parameter
is not annihilated by $b_0^-$, and the one-form,
sensitive to the phase of the local coordinates, is difficult
to define globally.
This time the state is annihilated by $b_0^-$ and the one-form
is phase independent. If we attempt to use the one-form
to cancel out the Weyl dependence of the integral of the two-form, the
relation ${\cal S}_1^{[2]} = d\, s_1^{[1]}$ holding globally,
 and Stokes theorem will
imply that we get a total result of zero. This concludes our plausibility
argument for the absence of a nontrivial CFT deformation induced by the
semirelative states ${\cal S}_1$ and ${\cal S}_2$. The analysis of
${\cal S}_3$ will be done shortly.

Consider now  the states that are
not annihilated by $b_0^-$.
These states, being outside the closed string state space,
do not correspond to linearized
solutions of the string equations of motion, and therefore do not
represent deformations of string backgrounds. One can ask if
such states can deform the CFT.
If this were the case we would have a problem with background
independence;  we would have two nearby conformal theories
giving rise to two string theories that
are not related by a shift of the string field. As we will explain
now we believe it is unlikely that  states which are not
annihilated by $b_0^-$ define
CFT deformations.

Consider the first state listed in \dsrstates.
The associated two-form is found to be given by
$${{\cal A}}_1^{[2]} =   -{1\over \sqrt 2} \, dz \wedge d\bar z\, \Bigl[
\partial X \, \bar \partial \,( \varphi - 2\sqrt 2\, i\theta ) \,\Bigr]\,,
\eqn\abss$$
where $\theta$ is the phase of the quantity $a(z_0,\bar z_0)$ appearing in the
definition of the local coordinate: $z-z_0 = a(z_0,\bar z_0) w + \O (w^2)$.
The deformation induced by this state
of the CFT partition function on a fixed
surface would be given
by integrating the above two-form over the complete surface.
Due to the non-zero Euler number, the phase
of the local coordinate cannot in general be defined globally throughout the
surface and the integral is not well defined.
It seems very unlikely that one can define a nontrivial CFT deformation
using the states in the absolute cohomology that are not annihilated by
$b_0^-$.

\section{The coupling constant in $D=2$ strings}

The ghost-dilaton,
always trivial in  absolute cohomology, becomes
trivial in semirelative cohomology for the background
defining $D=2$ string theory. Indeed, one readily verifies that
$$c\partial^2 c - \bar c \bar \partial^2 \bar c  = -\,{1
\over \sqrt{2}} \, \Bigl\{
Q \,,  \, c\partial \varphi - \bar c \bar\partial \varphi \, \Bigr\}\,.
\eqn\nowtriv$$
Note that $ c\partial \varphi - \bar c \bar\partial \varphi$ is fully legal;
it is a state in the standard semirelative complex.
Not only is the ghost-dilaton absent in
$D=2$ string theory, but now the last semirelative state
${\cal S}_3 = c\bar c
\partial X \bar \partial X$, is recognized to be trivial in
the extended complex.
This is because ${\cal S}_3$ is equivalent
to the ghost-dilaton in the extended complex.

Let us see explicitly why the string coupling is unobservable in
this background.
A change of coupling constant in
string field theory amounts to scaling
the string forms as
$$\bra{\Omega^{[d]g,n}} \to  \bra{\Omega^{[d]g,n}}
\Bigl( 1 - \e \,[2-2g-n\,]\Bigr)\,,\eqn\atlof$$
where   $d$ is the degree of the form.
On the other hand in $D=2$ strings
$$\bra{\Sigma_{g,n}\,} \sum_{i=1}^n \oint {dz\over 2\pi i} J^{(i)} (z) =
-2\sqrt{2} \,(2-2g)\,\bra{\Sigma_{g,n}\,}\,, \quad J^{(i)}(z) =
{\partial\varphi}^{(i)}(z)\,. \eqn\lianomali$$
Since this current has no ghost dependence, an identical relation
holds for string forms.
Since one can always add to the above right hand side a contribution
proportional to $n$ by adding a constant to the charge associated
to $J$, we see that the  deformation
\atlof\ can be implemented by a similarity transformation induced by $J$.
This means  that a homogeneous string field redefinition
changes the coupling constant making it  unobservable.
Indeed, the background we are considering,
called the linear dilaton vacuum, has a coordinate dependent
string coupling. The coupling is not observable since a shift
of coupling is equivalent to a translation along the $\varphi$ coordinate.

\refout
\end